\documentclass[fleqn,usenatbib]{mnras}
\usepackage[T1]{fontenc}

\usepackage{graphicx}	
\usepackage{flushend}

\newcommand{\rmaxs}{\ifmmode{R_{\rm{\sigma}}^{\rm{max}}}\else{$R_{\rm{\sigma}}^{\rm{max}}$}\fi}
\newcommand{\recirc}{\ifmmode{R_{\rm{e,c}}}\else{$R_{\rm{e,c}}$}\fi}
\newcommand{\re}{\ifmmode{R_{\rm{e}}}\else{$R_{\rm{e}}$}\fi}
\newcommand{\ret}{\ifmmode{R_{\rm{e/2}}}\else{$R_{\rm{e/2}}$}\fi}
\newcommand{\retwo}{\ifmmode{R_{\rm{e/2}}}\else{$2R_{\rm{e}}$}\fi}
\newcommand{\aee}{\ifmmode{a_{\rm{e}}}\else{$a_{\rm{e}}$}\fi}
\newcommand{\kpa}{\ifmmode{PA_{\rm{kin}}}\else{$PA_{\rm{kin}}$}\fi}
\newcommand{\ee}{\ifmmode{\varepsilon_{\rm{e}}}\else{$\varepsilon_{\rm{e}}$}\fi}
\newcommand{\epsi}{\ifmmode{\varepsilon_{\rm{intr}}}\else{$\varepsilon_{\rm{intr}}$}\fi}

\newcommand{\lr}{\ifmmode{\lambda_R}\else{$\lambda_{R}$}\fi}
\newcommand{\lre}{\ifmmode{\lambda_{R_{\rm{e}}}}\else{$\lambda_{R_{\rm{e}}}$}\fi}
\newcommand{\lret}{\ifmmode{\lambda_{R_{\rm{e/2}}}}\else{$\lambda_{R_{\rm{e/2}}}$}\fi}
\newcommand{\lretwo}{\ifmmode{\lambda_{2R_{\rm{e}}}}\else{$\lambda_{2R_{\rm{e}}}$}\fi}
\newcommand{\lreo}{\ifmmode{\lambda_{R_{\rm{e}}}^{\rm{\,obs}}}\else{$\lambda_{R_{\rm{e}}}^{\rm{\,obs}}$}\fi}
\newcommand{\lrei}{\ifmmode{\lambda_{R_{\rm{e}},\rm{intr}}}\else{$\lambda_{R_{\rm{e}},\rm{intr}}$}\fi}
\newcommand{\lreeo}{\ifmmode{\lambda_{\,R_{\rm{e}}}^{\rm{\,edge-on}}}\else{$\lambda_{\,R_{\rm{e}}}^{\rm{\,edge-on}}$}\fi}

\newcommand{\lreorig}{\ifmmode{\lambda_{\,R_{\rm{e}}}^{\rm{\,orig}}}\else{$\lambda_{\,R_{\rm{e}}}^{\rm{\,orig}}$}\fi}
\newcommand{\lrerep}{\ifmmode{\lambda_{\,R_{\rm{e}}}^{\rm{\,rep}}}\else{$\lambda_{\,R_{\rm{e}}}^{\rm{\,rep}}$}\fi}

\newcommand{\flr}{\ifmmode{f_{\lambda_{\,R}}}\else{$f_{\lambda_{\,R_{\rm{e}}}}$}\fi}
\newcommand{\flre}{\ifmmode{f_{\lambda_{\,R_{\rm{e}}}}}\else{$f_{\lambda_{\,R_{\rm{e}}}}$}\fi}
\newcommand{\flret}{\ifmmode{f_{\lambda_{\,R_{\rm{e/2}}}}}\else{$f_{\lambda_{\,R_{\rm{e/2}}}}$}\fi}

\newcommand{\vs}{\ifmmode{V / \sigma}\else{$V / \sigma$}\fi}
\newcommand{\vse}{\ifmmode{(V / \sigma)_{\rm{e}}}\else{$(V / \sigma)_{\rm{e}}$}\fi}
\newcommand{\vset}{\ifmmode{(V / \sigma)_{\rm{e/2}}}\else{$(V / \sigma)_{\rm{e/2}}$}\fi}
\newcommand{\vsetwo}{\ifmmode{(V / \sigma)_{\rm{2e}}}\else{$(V / \sigma)_{\rm{2e}}$}\fi}
\newcommand{\vseo}{\ifmmode{(V / \sigma)_{\rm{\,e}}^{\rm{\,obs}}}\else{$(V / \sigma)_{\rm{\,e}}^{\rm{\,obs}}$}\fi}
\newcommand{\vsei}{\ifmmode{(V / \sigma)_{\rm{\,e}}^{\rm{\,intr}}}\else{$(V / \sigma)_{\rm{\,e}}^{\rm{\,intr}}$}\fi}

\newcommand{\vobs}{\ifmmode{V_{\rm{obs}}}\else{$V_{\rm{obs}}$}\fi}
\newcommand{\sobs}{\ifmmode{\sigma_{\rm{obs}}}\else{$\sigma_{\rm{obs}}$}\fi}

\newcommand{\vseorig}{\ifmmode{(V / \sigma)_{{\rm{e}}}^{\rm{\,orig}}}\else{$(V / \sigma)_{{\rm{e}}}^{\rm{\,orig}}$}\fi}
\newcommand{\vserep}{\ifmmode{(V / \sigma)_{{\rm{e}}}^{\rm{\,rep}}}\else{$(V / \sigma)_{{\rm{e}}}^{\rm{\,rep}}$}\fi}

\newcommand{\se}{\ifmmode{\sigma_{\rm{e}}}\else{$\sigma_{\rm{e}}$}\fi}

\newcommand{\kms}{\ifmmode{\,\rm{km}\, \rm{s}^{-1}}\else{$\,$km$\,$s$^{-1}$}\fi}
\newcommand{\msun}{\ifmmode{~\rm{M}_{\odot}}\else{M$_{\odot}$}\fi}
\newcommand{\mstar}{\ifmmode{M_{\star}}\else{$M_{\star}$}\fi}
\newcommand{\logm}{\ifmmode{\log(M_{\star}/M_{\odot})}\else{$\log(M_{\star}/M_{\odot})$}\fi}
\newcommand{\loghm}{\ifmmode{\log(M_{\rm{halo}}/M_{\odot})}\else{$\log(M_{\rm{halo}}/M_{\odot})$}\fi}
\newcommand{\mm}{\ifmmode{M_{\star}/M_{\odot}}\else{$M_{\star}/M_{\odot}$}\fi}

\newcommand{\ea}{\textsc{eagle}}

\newcommand{\hy}{\textsc{hydrangea}}

\newcommand{\at}{\ifmmode{\rm{ATLAS}^{\rm{3D}}}\else{ATLAS$^{\rm{3D}}$}\fi}

\newcommand{\fsr}{\ifmmode{f_{\rm{SR}}}\else{$f_{\rm{SR}}$}\fi}
\newcommand{\sfivelong}{\ifmmode{\log_{10} \Sigma_{5, 1000, -18.5}}\else{$\log_{10} \Sigma_{5, 1000, -18.5}$}\fi}
\newcommand{\sfive}{\ifmmode{\log \Sigma_{5}}\else{$\log \Sigma_{5}$}\fi}

\newcommand{\sfiveu}{\ifmmode{\log ( \Sigma_{5}/\rm{Mpc}^{-2})}\else{$\log (\Sigma_{5}/\rm{Mpc}^{-2})$}\fi}
\newcommand{\sigpsf}{\ifmmode{\sigma_{\rm{PSF}}}\else{$\sigma_{\rm{PSF}}$}\fi}
\newcommand{\sigpsfre}{\ifmmode{\sigma_{\rm{PSF}}/R_{\rm{e}}}\else{$\sigma_{\rm{PSF}}/R_{\rm{e}}$}\fi}

\newcommand{\mk}{\ifmmode{\overline{k_5 / k_1}}\else{$\overline{k_5 / k_1}$}\fi}

\usepackage{newtxtext,newtxmath}

\title[Galaxy Dynamics and Environment]{The SAMI Galaxy Survey: 
Mass and Environment as Independent Drivers of Galaxy Dynamics}

\author[Jesse van de Sande]{Jesse van de Sande$^{1,2},$
Scott M. Croom$^{1,2}, $
Joss Bland-Hawthorn$^{1,2}, $
Luca Cortese$^{3,2}, $
Nicholas Scott$^{1,2}, $
\newauthor
Claudia D.P. Lagos$^{3,2},$
Francesco D'Eugenio$^{4},$
Julia J. Bryant$^{1,2,5}, $
Sarah Brough$^{6,2}, $
Barbara Catinella$^{3,2}, $
\newauthor
Caroline Foster$^{1,2}, $
Brent Groves$^{3,7,2}$
Katherine E. Harborne$^{3,2},$
\'Angel R. L\'opez-S\'anchez$^{8,9,10,2},$
\newauthor
Richard McDermid$^{9,10,2},$
Anne Medling$^{11,2},$
Matt S. Owers$^{9,10,2}, $
Samuel N. Richards$^{1}, $
Sarah M. Sweet$^{12,2} $
\newauthor
and Sam P. Vaughan$^{1,2}$
\\
\\
$^{1}${Sydney Institute for Astronomy, School of Physics, A28, The University of Sydney, NSW, 2006, Australia}\\
$^{2}${ARC Centre of Excellence for All Sky Astrophysics in 3 Dimensions (ASTRO 3D), Australia}\\
$^{3}${International Centre for Radio Astronomy Research, University of Western Australia, 35 Stirling Highway, Crawley WA 6009, Australia}\\
$^{4}${Sterrenkundig Observatorium, Universiteit Gent, Krijgslaan 281 S9, B-9000 Gent, Belgium} \\
$^{5}${The University of Sydney, NSW 2006, Australia; Australian Astronomical Optics}\\
$^{6}${School of Physics, University of New South Wales, NSW 2052, Australia}\\
$^{7}${Research School of Astronomy and Astrophysics, Australian National University, Canberra, ACT 2611, Australia}\\
$^{8}$ {Australian Astronomical Optics, Macquarie University, 105 Delhi Rd, North Ryde, NSW 2113, Australia}\\
$^{9}${Department of Physics and Astronomy, Macquarie University, NSW, 2109, Australia}\\
$^{10}${Astronomy, Astrophysics and Astrophotonics Research Centre, Macquarie University, Sydney, NSW 2109, Australia}\\
$^{11}${Ritter Astrophysical Research Center University of Toledo Toledo, OH 43606, USA}\\
$^{12}${School of Mathematics and Physics, University of Queensland, Brisbane, QLD 4072, Australia}
}

\date{Accepted XXX. Received YYY; in original form ZZZ}

\pubyear{2019}

\begin{document}
\label{firstpage}
\pagerange{\pageref{firstpage}--\pageref{lastpage}}
\maketitle


\begin{abstract}
The kinematic morphology-density relation of galaxies is normally attributed to a changing distribution of galaxy stellar masses with the local environment. However, earlier studies were largely focused on slow rotators; the dynamical properties of the overall population in relation to environment have received less attention.
We use the SAMI Galaxy Survey to investigate the dynamical properties of $\sim$1800 early and late-type galaxies with $\log(M_{\star}/M_{\odot})>9.5$ as a function of mean environmental overdensity ($\Sigma_{5}$) and their rank within a group or cluster. 
{By classifying galaxies into fast and slow rotators, at fixed stellar mass above $\log(M_{\star}/M_{\odot})>10.5$, we detect a higher fraction ($\sim3.4\sigma$) of slow rotators for group and cluster centrals and satellites as compared to isolated-central galaxies. We find similar results when using $\Sigma_{5}$ as a tracer for environment.}
Focusing on the fast-rotator population, we also detect a significant correlation between galaxy kinematics and their stellar mass as well as the environment they are in. Specifically, by using inclination-corrected or intrinsic $\lambda_{R_{\rm{e}}}$ values, we find that, at fixed mass, satellite galaxies on average have the lowest $\lambda_{\,R_{\rm{e}},\rm{intr}}$, isolated-central galaxies have the highest $\lambda_{\,R_{\rm{e}},\rm{intr}}$, and group and cluster centrals lie in between. Similarly, galaxies in high-density environments have lower mean $\lambda_{\,R_{\rm{e}},\rm{intr}}$ values as compared to galaxies at low environmental density. However, at fixed $\Sigma_{5}$, the mean $\lambda_{\,R_{\rm{e}},\rm{intr}}$ differences for low and high-mass galaxies are of similar magnitude as when varying $\Sigma_{5}$ {($\Delta \lambda_{\,R_{\rm{e}},\rm{intr}} \sim 0.05$, with $\sigma_{random}=0.025$, and $\sigma_{syst}<0.03$)}.
Our results demonstrate that after stellar mass, environment plays a significant role in the creation of slow rotators, while for fast rotators we also detect an independent, albeit smaller, impact of mass and environment on their kinematic properties.
\end{abstract}





\section{Introduction}
\label{sec:intro}

The $\Lambda$CDM (Lambda Cold Dark Matter) model predicts a strong dependence of galaxy properties on their environment \citep[e.g.][]{springel2005}, in particular because mergers are expected to play a vital role during the formation and/or evolution of almost every massive galaxy (e.g. \citealt{whiterees1978}). As interactions are more frequent for galaxies in groups as compared to isolated galaxies, a correlation between large-scale environment (groups and clusters) and galaxy properties is expected. Clear observational evidence for this paradigm comes from the morphology-density relation \citep{oemler1974,davis1976,dressler1980}: towards higher density environmental structures (e.g. clusters), the fraction of early-type galaxies increases, whereas the fraction of late-type galaxies decreases. Using a sample of nearly 50,000 galaxies from the Sloan Digital Sky Survey, \citet[SDSS; ][]{kauffmann2004} showed that the morphology-density relation can be partly explained by galaxies becoming more massive in high-density environments. However, \citet{bamford2009} showed that, at fixed stellar mass, the fraction of early-type galaxies is still higher in high-density environment, and \citet{peng2010} demonstrated the independent impact of mass and environment on the fraction of red galaxies.

Traditionally, galaxies are classified according to their morphological properties. But, in an era of large integral field spectroscopic (IFS) surveys, galaxies can now be classified according to their stellar kinematic properties. Specifically, by using a combination of observed ellipticity $\varepsilon$ and the ratio of ordered to random stellar motion, \vs, or a proxy for the spin parameter, \lr\ \citep[e.g.][]{cappellari2007,emsellem2007,emsellem2011,cappellari2016}, galaxies with high \lr\ are classified as fast rotators (FRs), whereas galaxies below a certain \lr\ and $\varepsilon$ threshold are labelled slow rotators (SRs). In the present-day Universe, the majority of early-type galaxies ($>85$ per cent) are FRs consistent with being axisymmetric, rotating oblate spheroids \citep{emsellem2011}, and only a minor fraction ($<15$ per cent) of galaxies are SRs with complex dynamical structures \citep[for a review, see][]{cappellari2016}. 

A link between the stellar dynamical properties of galaxies and environment is also suggested. \citet{cappellari2011b} show that the fraction of SRs (\fsr) is higher by a factor of two in the densest areas of the Virgo cluster as compared to lower-density environments. Within massive groups or clusters, this kinematic morphology-density relation for early-type galaxies has been reported by several other studies \citep{deugenio2013,houghton2013,scott2014,fogarty2014}. However, there appears to be no dependence of the fraction of SRs on the total mass of the group or cluster; on average, \fsr\ is also approximately 15 per cent in these environments \citep{brough2017} and has a strong dependence on galaxy luminosity or stellar mass \citep{emsellem2011,veale2017a,vandesande2017b,brough2017}. Recent results now show that galaxy stellar mass plays a more dominant role in changing the fraction of SRs than environment does \citep{brough2017,veale2017b,greene2018}, and that the projected environmental density relative to the peak density of groups or cluster is more fundamental than the absolute number density in impacting the fraction of slow rotators \citep{graham2019a}.

Using IFS-like ``observations'' of synthetic galaxies from the \ea\ and \hy\ cosmological hydrodynamic observations \citep{schaye2015, bahe2017}, \citet{lagos2018b} confirm the primary or strongest dependence of the \fsr\ on stellar mass, but find a weak, secondary dependence on environment \citep[see also][]{choi2018}. They find that at fixed stellar mass ($\logm\sim11.25$), satellite galaxies ($\fsr\sim0.28$) are less likely to be SRs than centrals ($\fsr\sim0.45$). Observationally, the fraction of FRs and SRs for both satellites and centrals has been explored in \citet{greene2018}, but they did not detect a significant difference at fixed stellar mass.

Most previous studies primarily focused on investigating the fraction of SRs in different environments. Therefore, the samples used in these studies have consisted solely of early-type galaxies, for two main reasons. First, the pioneering SAURON \citep{dezeeuw2002} and \at\ \citep{cappellari2011a} surveys were morphologically selected to consist of early-type galaxies only, and many consecutive studies, aimed at further investigating the kinematic morphology-density relation, followed this selection. Secondly, and perhaps more importantly, by construction it is nearly impossible for a late-type galaxy to be kinematically classified as an SR. Any galaxy with a clear detectable disk with spiral arms will have a \lr\ value that is too high to fall within the SR selection area. An exception is a face-on disk with low observed velocity and velocity dispersion. Including late-type galaxies in a sample will typically only lower the \fsr. 

In order to better understand if and how environment changes the dynamical properties of \emph{all} galaxy types, we need to use a sample that contains both early- \emph{and} late-type galaxies. The classic morphology-density relation shows a decreasing fraction of spiral galaxies and an increasing fraction of S0s towards denser environments. Therefore, by including late-type galaxies we can take a broader approach and investigate the full extent of distributions in the $\lr-\varepsilon$ plane.
However, environment has already been shown to have a smaller impact than stellar mass on the kinematic properties of galaxies \citep[e.g.][]{wang2020}. Thus, we also require a statistically significant sample of galaxies to control for stellar mass when studying the average kinematic properties across a range of environments. 

This type of study has only recently become possible with the introduction of new multi-object IFS surveys such as the SAMI Galaxy Survey \citep[Sydney-AAO Multi-object Integral field spectrograph; N $\sim3000$; ][]{croom2012} and the SDSS-IV MaNGA Survey \citep[Sloan Digital Sky Survey Data; Mapping Nearby Galaxies at APO; N $\sim10000$; ][]{bundy2015}. These surveys now allow for resolved kinematic measurements of thousands of galaxies across the full Hubble morphological sequence with a wide range in stellar masses and environment.

Here we investigate the impact of environment on the kinematic properties of galaxies using a proxy for the spin parameter \lr. The main goal of this paper is to separate the impact of stellar mass and environment (traced by different metrics) on the kinematic properties of galaxies. The paper is organized as follows: Section \ref{sec:obs_data} presents the data from the observations. In Section \ref{sec:dyn_prop} we first follow the approach of previous studies by measuring and analysing the fraction of SRs, and then explore an alternative method for tracing the dynamical differences of all galaxies. We discuss the implication of our results in Section \ref{sec:discussion} and summarise and conclude in Section \ref{sec:conclusion}. Throughout the paper we assume a $\Lambda$CDM cosmology with $\Omega_\mathrm{m}$=0.3, $\Omega_{\Lambda}=0.7$, and $H_{0}=70$ km s$^{-1}$ Mpc$^{-1}$.

\section{Observational Data and Measurements}
\label{sec:obs_data}

\subsection{SAMI Galaxy Survey}
\label{subsec:sgs}

We use a sample of nearby ($z \lesssim 0.1$) galaxies with spatially resolved spectroscopic observations from the SAMI Galaxy Survey \citep{bryant2015}, conducted with the SAMI instrument \citep{croom2012}. SAMI is mounted at the prime focus of the 3.9m Anglo Australian Telescope (AAT). This multi-object IFS can simultaneously observe 12 galaxies with imaging fibre bundles, or \textit{hexabundles} \citep{blandhawthorn2011,bryant2011,bryant2012a,bryant2014}, that are each manufactured from 61 individual fibres with 1\farcs6 angle on sky. Each hexabundle is deployable over a 1$^\circ$ diameter field-of-view, covers a $\sim15$ arcsec diameter region on the sky, and has a maximal filling factor of 75 per cent. All 819 fibres, including 26 individual sky fibres, are fed to the AAOmega dual-beamed spectrograph \citep{saunders2004, smith2004, sharp2006}. 

The SAMI Galaxy Survey finished observations in May 2018 and has observed over $\sim$3000 galaxies covering a broad range in galaxy stellar mass ($M_{\star}= 10^{8}-10^{12}$\msun) and galaxy environment (field, groups, and clusters). We use the final data release DR3 \citep{croom2021a} that contains 3068 unique galaxies. The adopted redshift range of the survey ($0.004<z<0.095$) results in a spatial resolution of 1.6 kpc per fibre at $z=0.05$. The Galaxy and Mass Assembly \citep[GAMA;][]{driver2011} Survey was used to select field and group targets in four volume-limited galaxy samples derived from cuts in stellar mass in the GAMA G09, G12 and G15 regions. GAMA is a major campaign that combined a large spectroscopic survey of $\sim$300,000 galaxies carried out using the AAOmega multi-object spectrograph on the AAT, with a large multi-wavelength photometric data set. Additionally, we selected cluster targets from eight high-density cluster regions sampled within radius $R_{200}$ with the same stellar mass limit as used for the GAMA fields \citep{owers2017}. 

The SAMI Galaxy Survey employed both the blue (3750-5750\AA) and red (6300-7400\AA) arms of the AAOmega spectrograph, using the 580V and 1000R grating, respectively. The resulting spectral resolution is R$_{\rm{blue}}\sim 1810$ at 4800\AA, and R$_{\rm{red}}\sim4260$ at 6850\AA\ \citep{scott2018}. In order to cover gaps between fibres and to create data cubes with 0\farcs5 spaxel size, all observations are carried out using a six- to seven-position dither pattern \citep{sharp2015,allen2015}. 
Reduced data-cubes and stellar kinematic data products for all galaxies are available on: \url{https://datacentral.org.au}, as part of the first, second, and third SAMI Galaxy Survey data release \citep{green2018,scott2018,croom2021a}.

\subsection{Ancillary Data}
\label{subsec:ancil}

We use the aperture-matched $g$ and $i$ photometry from the GAMA catalogue \citep{hill2011,liske2015}, measured from reprocessed SDSS Data Release Seven \citep{york2000, kelvin2012}, and for the clusters we use both SDSS (DR9) and VLT Survey Telescope ATLAS imaging data  \citep{shanks2013,owers2017}, to derive $g-i$ colours. Stellar masses are derived from the rest-frame i-band absolute magnitude and $g-i$ colour \citep{bryant2015} by using the colour-mass relation following the method of \citet{taylor2011}. For the stellar mass estimates, a \citet{chabrier2003} stellar initial mass function (IMF) and exponentially declining star formation histories are assumed. For more details see \citet{bryant2015}. 

Effective radii, ellipticities, and positions angles measurements are described in \citet{deugenio2021}, derived using the Multi-Gaussian Expansion \citep[MGE;][]{emsellem1994,cappellari2002} technique and the code from \citet{scott2009,scott2013} on imaging from the GAMA-SDSS \citep{driver2011}, SDSS \citep{york2000}, and VST/ATLAS \citep{shanks2013,owers2017}. \re\ is defined as the semi-major axis effective radius, and the ellipticity of the galaxy within one effective radius is defined as $\varepsilon_{\rm{e}}$, measured from the best-fitting MGE model. 

Galaxies' visual morphologies are determined using classifications that are based on the SDSS and VST colour images; galaxies are divided according to their shape, presence of spiral arms and/or signs of star formation \citep[for more details see][]{cortese2016}. A total of 1289 galaxies are classified as Elliptical or S0 (42.0 per cent early types; SAMI $0\leq Mtype<2$), whereas 1631 are labelled Early-Spiral (Sa/Sb), Late-Spiral (Sc/Sc) or Irregular (53.2 per cent late types; SAMI $Mtype\geq2$). For 148 galaxies (4.8 per cent) out of the total 3068 galaxies in the observed SAMI sample, no conclusive morphology could be determined.

\subsection{Density and Environment Estimates}
\label{subsec:dens_est}

We determine the local environment of galaxies using a nearest-neighbour density estimate to probe the underlying density field, with the assumption that galaxies with closer neighbours are also in denser environments \citep{muldrew2012}. Here we combine the GAMA regions of the SAMI Galaxy Survey with the SAMI Cluster Survey regions, but we note that the kinematic morphology-density relation using only the SAMI cluster sample is also presented in \citet{brough2017}. To estimate the local environment, we use the method as outlined in \citet{brough2013,brough2017} and \citet{croom2021a}. The local surface density is defined as $\Sigma_{\rm{N,Vlim,Mlim}}$. Here, $\Sigma$ is the surface density derived using the projected comoving distance to the Nth nearest neighbour with a velocity limit $\pm$V$_{\rm{lim}}$ \kms, and a volume-limited density-defining population with an absolute magnitude \mbox{M$_r$ < $M_{\rm{lim}}-Q_{z}$}. We use $Q_z=1.03$ which is defined as the expected evolution of $M_r$ as a function of redshift, $z$, \citep{loveday2015}. We adopt \sfivelong: $N=5$ for the fifth nearest neighbour, V$_{\rm{lim}}$=1000\kms, and $M_{\rm{lim}}$=-18.5 mag, abbreviated to \sfive. We refer to \citet{brough2017} for a study on the impact of using different limits to derive the surface density. Galaxies near the edge of the GAMA and cluster regions suffer from larger uncertainties in the overdensity measurements due to the lack of deep, high-completeness spectroscopy beyond the edge of the survey. These galaxies are excluded from the sample when analysing trends over bins of \sfive\ ($\sim17$ per cent of the total stellar kinematic sample in the GAMA region, $<1$ per cent for cluster targets; see \citealt{croom2021a}).

Besides the local environmental parameter \sfive, we also use the GAMA galaxy group catalogue (G3Cv1, v010) from \citet{robotham2011} to identify group central and satellite galaxies. G3Cv1 is a parametric approach to recover the underlying group statistics using a friends-of-friends based algorithm. The code uses both the radial (inferred from galaxy redshifts) and projected separations to disentangle projection effects and has been designed to be robust against the effects of outliers and linking errors. We classify galaxies not identified in the GAMA group catalogue as isolated. Note that in galaxy formation simulations, central galaxies are those that sit at the centre of the potential well, and hence our observational sample of isolated galaxies is likely to be mostly composed by centrals. However, we adopt a classification of "isolated-central" and "group/cluster centrals" in order to distinguish between central galaxies with no observed satellites in the field (i.e. isolated) and centrals of groups and clusters. Finally, we note that below $z<0.2$, the GAMA field G09 is underdense compared to G12 and G15, whereas at $z<0.1$ the GAMA fields are overall 15 per cent underdense with respect to SDSS DR7 \citep{driver2011}. 

Within the eight SAMI clusters, we follow the approach from \citet{santucci2020} and define the central galaxy as the most massive galaxy within a radius of 0.25$R_{200}$ using the cluster centres as defined in \citet{croom2021a}. All other cluster galaxies are defined as satellites. Note that for Abell 168, Abell 2399 and Abell 4038, the central galaxy is different from the brightest cluster galaxy as defined in \citet{owers2017}. The definition of the central galaxy in Abell 168 and Abel 2399 clusters is further complicated as there are multiple substructures due to undergoing cluster mergers. However, as this complication only impacts three cluster centrals out of a total 440 group and cluster centrals in our sample, classifying the central from a different substructure in these clusters does not significantly impact our results.

\subsection{Stellar Kinematic Measurements}
\label{subsec:stelkin_sami}

\subsubsection{Method}
\label{subsubsec:method}
The stellar kinematic measurements for the SAMI Galaxy Survey are described in detail in \citet{vandesande2017}. We summarise our method below. The penalized pixel fitting code \citep[pPXF;][]{cappellari2004,cappellari2017} is used and we assume a Gaussian line-of-sight velocity distribution (LOSVD). We convolve the spectral resolution of the red arm to match the instrumental resolution in the blue. Both blue and red spectra are then rebinned and combined onto a logarithmic wavelength scale with constant velocity spacing (57.9 \kms). From SAMI annular-binned spectra, we derive a set of radially-varying optimal templates using the MILES stellar library \citep{sanchezblazquez2006,falconbarroso2011}. For each individual spaxel, we allow \textsc{pPXF} to use the optimal templates from the annular bin in which the spaxel is located as well as the optimal templates from neighbouring annular bins. The uncertainties on the LOSVD parameters are estimated from 150 simulated spectra.


\begin{figure}
	\includegraphics[width=\linewidth]{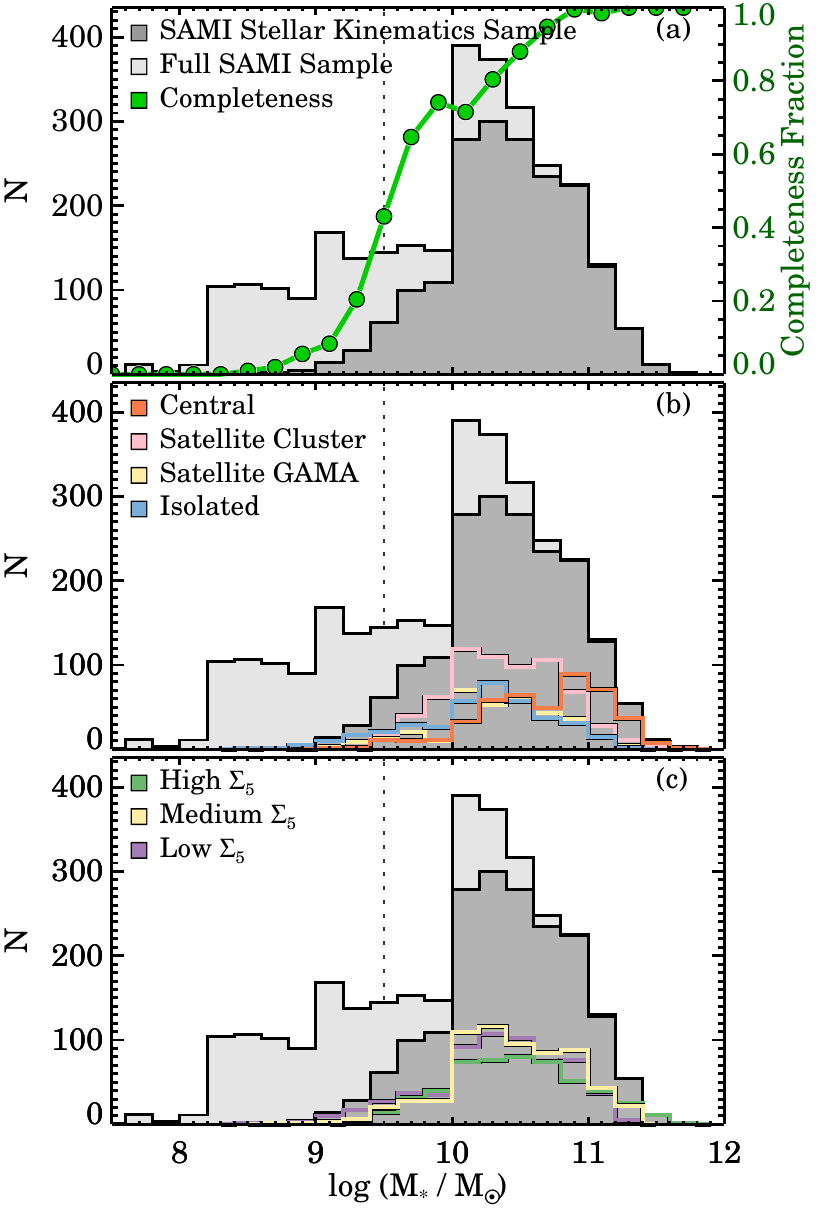}
    \caption{Stellar mass distribution of the various SAMI sub-samples. In Panel (a) we show the full SAMI sample and the stellar kinematic sample, and find that the stellar kinematic sample is biased towards high stellar mass, with 50 per cent completeness reached at $\logm\sim9.5$. The stellar kinematic sample split by galaxy group rank (Panel b) reveals a larger fraction of isolated-central galaxies towards lower stellar mass and a larger number of group and cluster central galaxies towards higher stellar mass, with group and cluster satellites in between. Based on three \sfive\ bins, in Panel (c) we detect no clear group environmental dependence as a function of stellar mass.}
    \label{fig:group_iso_central}
\end{figure}


\subsubsection{\lre -- A Proxy for the Spin Parameter}
\label{subsubsec:vs_lr}

Throughout the rest of this analysis, we quantify the dynamical properties of galaxies predominantly using the (\lr, $\varepsilon$) diagram \citep{emsellem2007}. \lre\ is a proxy for the spin parameter and is related to the average ratio of the velocity and velocity dispersion within one effective radius \vse\ \citep[see for example ][]{emsellem2011,vandesande2017b}. It quantifies the ratio of the ordered rotation and the random motion in a stellar system. Our adopted spin parameter is the proxy value \lr\ given by \citet{emsellem2007}:

\begin{flushleft}
\begin{equation}
\label{eq:lr}
\lambda_{R} = \frac{\langle R |V| \rangle }{\langle R \sqrt{V^2+\sigma^2} \rangle } = \frac{ \sum_{j=0}^{N_{\rm{spx}}} F_{j}R_{j}|V_{j}|}{ \sum_{j=0}^{N_{\rm{spx}}} F_{j}R_{j}\sqrt{V_j^2+\sigma_j^2}}.
\end{equation}
\end{flushleft}

\noindent The sum is taken over all spaxels $N_{\rm{spx}}$ within an ellipse with semi-major axis \re\ and axis ratio $b/a$. The subscript $j$ refers to the position of a spaxel within the ellipse, $F_{j}$ the flux of the $j^{th}$ spaxel, $V_{j}$ is the stellar velocity in \kms, $\sigma_{j}$ the velocity dispersion in \kms. $R_{j}$ is the semi-major axis of the ellipse on which spaxel $j$ lies, not the circular projected radius to the center as is used by e.g. \at\ \citep{emsellem2007}. We use the unbinned flux, velocity, and velocity dispersion maps as described in Section \ref{subsubsec:method}.  We use the input galaxy catalogue's R.A. and Dec. and WCS information from the cube headers, to determine a galaxy's centre. The systemic velocity is determined from 9 central spaxels ($1\farcs5 \times 1\farcs5$ box). We only use spaxels that meet the quality criteria for SAMI Galaxy Survey data as described in \citet{vandesande2017}: signal-to-noise (S/N) $>3$\AA$^{-1}$, \sobs $>$ FWHM$_{\rm{instr}}/2 \sim 35$\kms\ where the FWHM is the full-width at half-maximum, $V_{\rm{error}}<30$\kms \citep[Q$_1$ from][]{vandesande2017}, and $\sigma_{\rm{error}} < \sobs *0.1 + 25$\kms\ \citep[Q$_2$ from][]{vandesande2017}. 

We include a \lre\ seeing correction from \citet{harborne2020b}, optimised for the SAMI Galaxy Survey \citep{vandesande2021} applied to all galaxies, to create an unbiased \lre\ distribution. Seeing impacts galaxies with smaller angular sizes more severely. Combined with intrinsic differences in the physical sizes of both early and late-types and a redshift-dependent mass selection, the impact of seeing on IFS measurements can lead to a  morphologically biased \lre\ distribution. For more details on the seeing correction in relation to the selection of fast and slow rotators we refer to \citet{vandesande2021}.

Not all SAMI measurements have coverage out to one effective radius. Therefore, a \lr\ and \vs\ aperture correction is applied to all galaxies where the fill factor of good spaxels within one effective radius is less than 95 per cent, as outlined in \citet{vandesande2017b}. 
{The aperture correction method is based on the fact that for a large number of aperture measurements in the SAMI and \at\ data, a tight relation exists for \lr\ between different apertures, such that \lr\ increases as a function of radius $R$. This tight relation allows us to recover \lre\ with a mean uncertainty of 11 per cent when the aperture size is less than one \re. }
A total of 267 SAMI galaxies {in the final stellar kinematic sample} are aperture corrected (15 per cent{; 267/1766}). 

\subsubsection{\lrei\ -- Correcting for Inclination}
\label{subsubsec:lredgeon}

We derive edge-on or intrinsic \lre\ value from the observed \lre\ and \ee\ measurements combined with theoretical predictions from the tensor virial theorem that links velocity anisotropy, rotation and intrinsic shape \citep{binney2005}. The main assumption is that galaxies are simple rotating oblate axisymmetric spheroids with varying intrinsic shape and mild anisotropy $\beta_z = 0.7 \times \epsi$ \citep{cappellari2007}. 

While this is an oversimplification of the known complexities of galaxy structure and dynamics, {in particular for triaxial slow-rotators \citep[e.g.][]{vandenbosch2010,walsh2012,thomas2014}}, 
{we note that \citet{weijmans2014} and \citet{foster2017} demonstrate that fast-rotating galaxies are consistent with having oblate shapes, as derived from inverting the distributions of apparent ellipticities and the alignment or misalignment of the photometric and kinematic position angles. Specifically, they find that the observed intrinsic ellipticity distribution derived from \lre\ and \ee\ is similar to the intrinsic ellipticity distribution from the statistical inversion method.} 

Other methods for inclination correcting \lr\ exist and we compare several different methods in Appendix \ref{sec:app_eps_intr}, {from which we estimate that the median random uncertainty on \lrei\ is $\sim0.025$ with potential systematic uncertainties of \lrei<0.03.} Our adopted method starts with the relation between the observed \ee\ and intrinsic ellipticity \epsi\ as given by \citet{cappellari2016}:

\begin{equation}
\ee\ = 1 - \sqrt{1 + \varepsilon_{\rm intr}(\varepsilon_{\rm intr}-2)/\sin^2 i},
\end{equation}

\noindent where $i$ is the inclination. For different values of the inclination, the observed \vse\ can be calculated from the intrinsic \vs\ using:

\begin{equation}
\left(\frac{V}{\sigma}\right)_{\rm\! e}^{\rm\! obs} = 
\left(\frac{V}{\sigma}\right)_{\rm\! e}  \frac{\sin i}{\sqrt{1 - \delta\cos^2 i}}.
\end{equation}

\noindent Here $\delta$ is related to the anisotropy parameter $\beta_z$ such that $\delta\approx\beta_z=0.7\times\varepsilon_{\rm intr}$. The relation between \vs\ and \epsi\ for an edge-on view ($i=90$) is given by \citet{cappellari2007}:

\begin{equation}
\left(\frac{V}{\sigma}\right)_{\rm\! e}^{\rm\! intr} = \sqrt{ \frac{(0.09+0.1\ \epsi)\ \epsi}{1-\epsi}}.
\label{eq:vs_epsintr}
\end{equation}

\noindent We then use the relation between \vse\ and \lre\ as given by \cite{emsellem2011}:

\begin{equation}
    \lr = \frac{\kappa~(\vs)}{\sqrt{1+\kappa^2~(\vs)^2}}.
	\label{eq:lr_vs}
\end{equation}

\noindent For SAMI Galaxy Survey data, we adopt the empirical best-fit value of $\kappa=0.97$ from \citet{vandesande2017b}. In order to estimate the intrinsic \lre\ values for each individual galaxy, we first construct a large grid with different values for \epsi\ and inclination $i$. We then calculate the equivalent observed ellipticity and \lre\ values, and the nearest grid-point gives the best estimate of a galaxy's intrinsic ellipticity and inclination. For extremely rounds objects ($\ee<0.025$) we add a small value of 0.025 to \ee, to avoid the region where the model predictions become highly degenerate. This limits the estimated \lrei\ values for these galaxies to more conservative values, but we note that this has no impact on the key results as presented in Section \ref{subsec:lr_eo_environment}. The grid step size is set to $\Delta\epsi=0.001$ and $\Delta i=0.1^{\circ}$. With the recovered \epsi\ values we then calculate the intrinsic \vs\ using Eq.~\ref{eq:vs_epsintr} and convert these values to \lrei\ using Eq.~\ref{eq:lr_vs}. For galaxies that lie outside the range of the model prediction (i.e. below the magenta line in Fig.~\ref{fig:mass_ellip_lambdar}c), we assume that \lrei = \lre, {i.e. we do not apply an inclination correction but use the observed \lre}. 


\subsection{Stellar Kinematic Sample Properties}
\label{subsubsec:sample}

\subsubsection{Distribution of \lr, \ee, and stellar mass}
Of the 3068 SAMI kinematic maps in the GAMA and cluster regions, through visual inspection we flag and exclude 136 galaxies with irregular kinematic maps due to nearby objects or mergers that influence the stellar kinematics of the main object.
{We additionally excluded 763 galaxies where the S/N beyond R>2\farcs0 is insufficient to accurately measure the stellar kinematics. A further 261 galaxies are rejected because they are unresolved having \re<1\farcs5.}
Furthermore, we exclude another 40 galaxies where the ratio of the point-spread-function versus the effective radius of a galaxy is larger than $\sigpsfre>0.6$. This limit is chosen because of the relatively large impact of beam-smearing on \lre\ and \vse\ at these  $\sigpsfre$ values (see Appendix C of \citealt{harborne2020b} and \citealt{vandesande2021}). Lastly, for another 35 galaxies no reliable \lr\ aperture correction out to one \re\ could be derived (see Section \ref{subsubsec:vs_lr}). This brings the total sample of galaxies with kinematic measurements to 1833.


\begin{figure}
\begin{center}
	\includegraphics[width=0.90\linewidth]{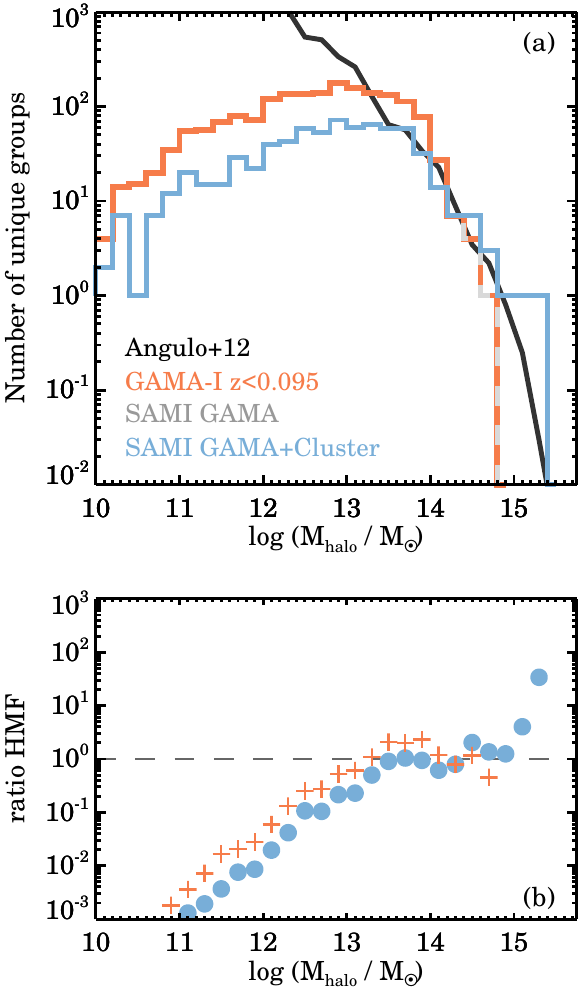}
    \caption{Distribution of group and cluster halo masses in the GAMA Survey (orange) and SAMI Galaxy Surveys (grey, blue) as compared to a theoretical halo mass function from \citet[][black]{angulo2012}. We include a group only once if multiple GAMA or SAMI galaxies are observed from the same group. There is a lower than predicted number of high-halo mass groups and clusters in the GAMA regions (orange \& dashed grey line; Panel a), whereas in the SAMI Galaxy Survey that includes the cluster regions we find an over-abundance above \loghm>15 (solid blue line; Panel a). Panel (b) shows the ratio of the GAMA and SAMI Galaxy Survey halo mass distributions over the theoretical halo mass distribution. Between $13<\loghm<15$ the SAMI Galaxy Survey closely matches the predicted halo mass distribution.}
    \label{fig:hm_distr}
\end{center}
\end{figure}


We note that a relatively large fraction of galaxies that are excluded here are galaxies with stellar mass less than $\logm<9.5$. Because the stellar kinematic completeness drops rapidly below 50 per cent at low stellar mass (see Fig.~\ref{fig:group_iso_central}), we do not use the remaining 67 galaxies below \logm<9.5 for the core analysis of this paper. The final number of galaxies from the SAMI Galaxy Survey with usable stellar velocity and stellar velocity dispersion maps above a stellar mass of \logm>9.5 is 1766; we dub this set of galaxies the "SAMI stellar kinematic sample".

The mass distribution of the SAMI stellar kinematic sample is compared to the full observed sample in Fig.~\ref{fig:group_iso_central}a. We find a clear drop in the total number of SAMI galaxies just below $\logm \sim 10$ caused by cosmic variance in the GAMA regions combined with the SAMI step-function selection \citep[see ][]{bryant2015}. The stellar kinematic sample is clearly biased towards higher stellar mass compared to the main sample, with few \lre\ measurements below a stellar mass of $\logm \sim 9$.
The different coloured lines in Fig.~\ref{fig:group_iso_central}b show the distribution of group or cluster central, satellite, and isolated central galaxies. At high stellar masses central galaxies dominate, whereas at low stellar masses isolated galaxies are more abundant. However, based on a comparison to the full SAMI sample, we do not find a significant bias in the stellar kinematic sample towards central, satellite, or isolated galaxies.
{Specifically, above \logm>9.5, the galaxies that are excluded from the sample have the same ratio of central, satellite, or isolated galaxies as compared to the stellar kinematic sample. Below \logm<9.5, as expected isolated galaxies are most abundant, followed by satellite galaxies in GAMA, and  group centrals. Note that because of the redshift range of the clusters, no cluster galaxies are observed below \logm<9.5.}

Fig.~\ref{fig:group_iso_central}c shows the stellar kinematic sample split in three bins of mean overdensity as described in Section \ref{subsec:fsr}. At fixed stellar mass, we find an almost equal number of galaxies in all three environment bins, but at the highest stellar masses ($\logm>11.2$) we find more galaxies in the high-density bin. {When we compare the distribution of environments in the stellar kinematic sample to the full SAMI sample, we find that above \logm>9.5 excluded galaxies have the same ratio of galaxies in the three environment bins, whereas below \logm<9.5 galaxies in low-density environments dominate.}


\begin{figure}
	\includegraphics[width=0.95\linewidth]{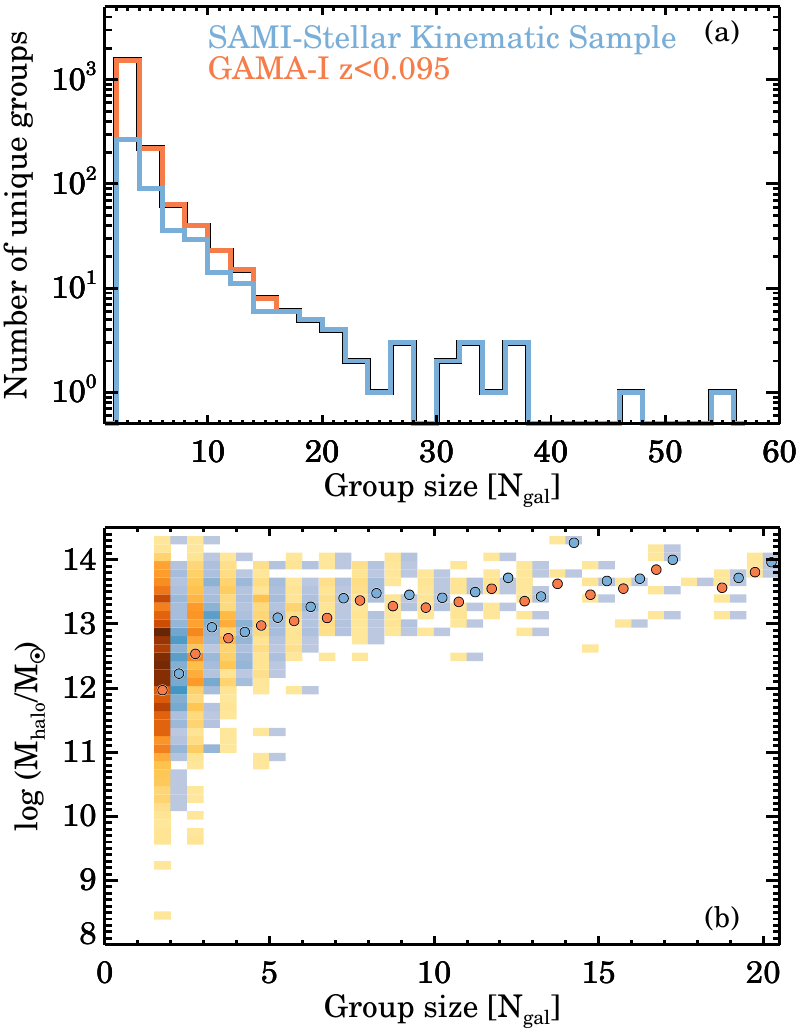}
    \caption{Group properties of GAMA and SAMI galaxies at $z<0.095$. From the normalised distribution of unique groups as a function of group size (Panel a) we find that galaxies in the SAMI kinematic sample reside in groups with a higher occupancy. Panel (b) compares group halo mass and number of group members for GAMA (orange squares, where darker grey means higher density of galaxies) and SAMI (blue squares, offset by $N_{\rm{gal+0.5}}$); orange and blue circles indicate the median halo mass for a given group occupancy for the GAMA and SAMI sample, respectively.}
    \label{fig:group_prop}
\end{figure}



\begin{figure*}
	\includegraphics[width=1.0\linewidth]{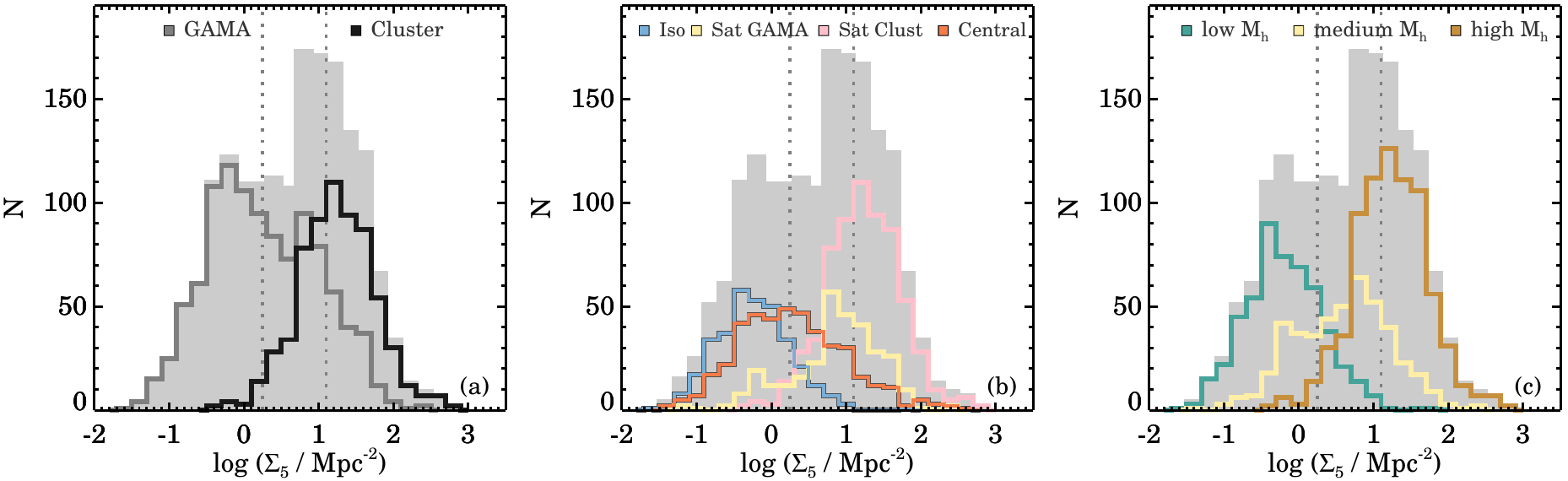}
	    \vspace{-0.5cm} 
    \caption{Distribution of mean environmental overdensity \sfive\ split into GAMA versus Cluster galaxies (Panel a), group rank (Panel b), and halo mass (Panel c). The dotted lines show the separation of the three bins in \sfive as described in the text, whereas we adopt the following three limits for the low, medium, and high halo mass bins: \loghm<12.5, 12.5<\loghm<14, \loghm>14. In low-density environments we predominantly find galaxies in isolation, being group centrals, or in low-mass haloes, whereas high-density environments harbour the largest fraction of satellite galaxies and galaxies in high-mass haloes. This high fraction of satellite galaxies is a natural consequence of the fact that massive groups and clusters only contain one central galaxy, yet many satellites.}
    \label{fig:group_vs_env}
\end{figure*}


\subsubsection{Halo mass Distribution}
\label{subsubsec:hm_distr}

The SAMI Galaxy Survey observed a combination of targets drawn from the volume-limited GAMA survey and 8 cluster regions. As the GAMA regions lack high over-density regions with halo mass greater than $\loghm\sim14.5$, galaxies within cluster regions were added to fill this density gap. To investigate whether the combined sample provides a representative halo mass distribution, we calculate the total survey volume, using the stepped series of stellar mass limits as a function of redshift from which the SAMI Galaxy Survey targets were selected \citep[see][]{bryant2015}. For each volume, we then calculate the observed halo mass function using the group halo masses from the GAMA-G3Cv1 catalogue \citet[v010][]{robotham2011} and for the clusters using the data from \citet{owers2017}. Similarly, for each volume, we also derive the predicted halo mass function from \citet{angulo2012} using \textsc{HMFcalc}: An Online Tool for Calculating Dark Matter Halo Mass Functions \citep{murray2013}. With that halo mass function, we can then obtain a probability of finding a cluster galaxy within the SAMI-GAMA volume. 

Fig.\ \ref{fig:hm_distr} presents the GAMA halo mass distribution, the SAMI Galaxy Survey halo mass distribution with and without the cluster sample, as well as the  \citet{angulo2012} halo mass distribution function. For the GAMA regions, the observed halo mass function under-samples the theoretical prediction towards the massive end (for haloes above \loghm>14.7, Fig.~\ref{fig:hm_distr}a), although note that the predicted number of haloes in each bin is $\leq2$. There is also a clear lack of low-mass haloes (\loghm<13) that is expected given the depth and sensitivity of GAMA \cite[e.g.][]{robotham2011}. The SAMI Galaxy Survey halo mass distribution follows the one from GAMA but becomes incomplete towards lower mass haloes (\loghm<14), as SAMI does not target galaxies to the faint limit of GAMA.

With the cluster sample included (Fig.\ \ref{fig:hm_distr}a), the SAMI Galaxy Survey follows the predicted halo mass distribution function closely up to $\loghm\sim15$, but for higher halo masses the observed sample has a higher than predicted number of haloes. As this over-abundance of galaxies in high-density environments could impact our results when comparing the dynamical properties as a function of stellar mass and environment, we calculate a halo mass weighting factor for each galaxy. Depending on the stellar mass of galaxy, we determine the ratio of the predicted halo mass function as compared to the observed (e.g. Fig.~\ref{fig:hm_distr}b), and use this as the weighting factor. For example, a galaxy in the most massive SAMI Galaxy Survey cluster (Abell 85) will receive a weight of 1/38, whereas a typical galaxy in the GAMA region will have a weight of 1. These weights are included throughout the rest of the analysis that follows, i.e. we calculate weighted fractions, means, and distributions.

Even though the observed and theoretical halo mass functions strongly diverge below $\loghm<13$, we chose not to apply a correction at these low halo masses as the uncertainties on the halo mass estimates for galaxies in groups of two are large. Instead, we only apply the halo mass correction factor to galaxies that reside in haloes with $\loghm>14.5$ (37.5 per cent, 662/1766), where the SAMI Galaxy Survey has an over-abundance of galaxies in high-density environments.

\subsubsection{Group Statistics and Environment Density Distribution}
We show the group statistics of the GAMA sample at $z<0.095$ and the SAMI stellar kinematic sample in Fig.~\ref{fig:group_prop}. As a function of galaxy group size (Panel a), we find that the largest groups ($N>10$) are nearly all groups have at least one target observed with SAMI, due to SAMI having brighter selection limits than GAMA. However, towards smaller groups, the incompleteness rises which indicates that not every group in GAMA has a galaxy observed with SAMI. In Fig.~\ref{fig:group_prop}b, we check whether this biases the median group mass at fixed group membership, but detect no significant difference between the median SAMI stellar kinematic and the GAMA samples.

Fig.~\ref{fig:group_vs_env} shows the mean environmental overdensity \sfive\ distributions comparing different environmental metrics. We first divide the sample into SAMI galaxies observed in the GAMA versus the cluster regions (Fig.~\ref{fig:group_vs_env}a). As the GAMA regions contain few groups with halo masses above $10^{14}\msun$, it is no surprise that the high density environments are dominated by galaxies from the cluster sample. However, around $\sfiveu=1$ there is a roughly equal number of GAMA and cluster galaxies. For the analysis that follows, we use three different environmental density bins: low, medium, and high. We choose the limit at $\sfiveu=0.25$ and $\sfiveu=1.1$ to get roughly equal numbers of galaxies in each bin. The lower limit is just to the right of the peak of the GAMA \sfive\ distribution, whereas the higher limit is at the peak of the cluster \sfive\ distribution. The three bins contain 583, 587, and 555 galaxies from low to high \sfive, respectively.

We show the different distributions of group and cluster central, satellite (GAMA groups and SAMI clusters), and isolated central galaxies in Fig.~\ref{fig:group_vs_env}b. At high overdensity the majority of galaxies are classified as satellites, which is expected as groups get more massive and only one galaxy in the group or cluster can be a central. While the median \sfive\ of isolated galaxies is lower than centrals and satellites, the distribution of isolated galaxies stretches well into the medium density bin. Similarly, we find central and satellite galaxies down to low \sfive. The relatively large overlap of these different environmental classifications can be understood from the fact that the group catalogue starts with N=2, whereas the mean environmental density is determined from the fifth nearest neighbour. Hence, groups consisting of a pair of galaxies can be at low density. Furthermore, groups of $N=2$ count satellites that are at the GAMA magnitude limit ($r$=19.8 mag), which is fainter than the density-defining population that is used in measuring \sfive. The complementary nature of the two definitions is the main motivation for using both environmental classifications side-by-side throughout the paper.

The distribution of different halo masses as a function of mean overdensity is shown in Fig.~\ref{fig:group_vs_env}. The lower limit of \loghm<12.5 typically traces galaxies in isolated haloes \citep[][ although this might be a completeness artefact of group finders]{yang2005}, whereas the higher limit of \loghm>14 is a common cutoff between large groups and clusters (e.g. Fornax versus Virgo). The three bins contain 523, 469, and 773 galaxies from low to high \loghm, respectively. There is good agreement between halo mass and \sfive, albeit with considerable overlap. As halo mass and the \sfive\ measurements trace the same overdensities, we chose not to present the results for halo masses in the core of this paper, but instead show the halo mass results in Appendix \ref{sec:halo_mass}.


\begin{figure*}
	\includegraphics[width=\linewidth]{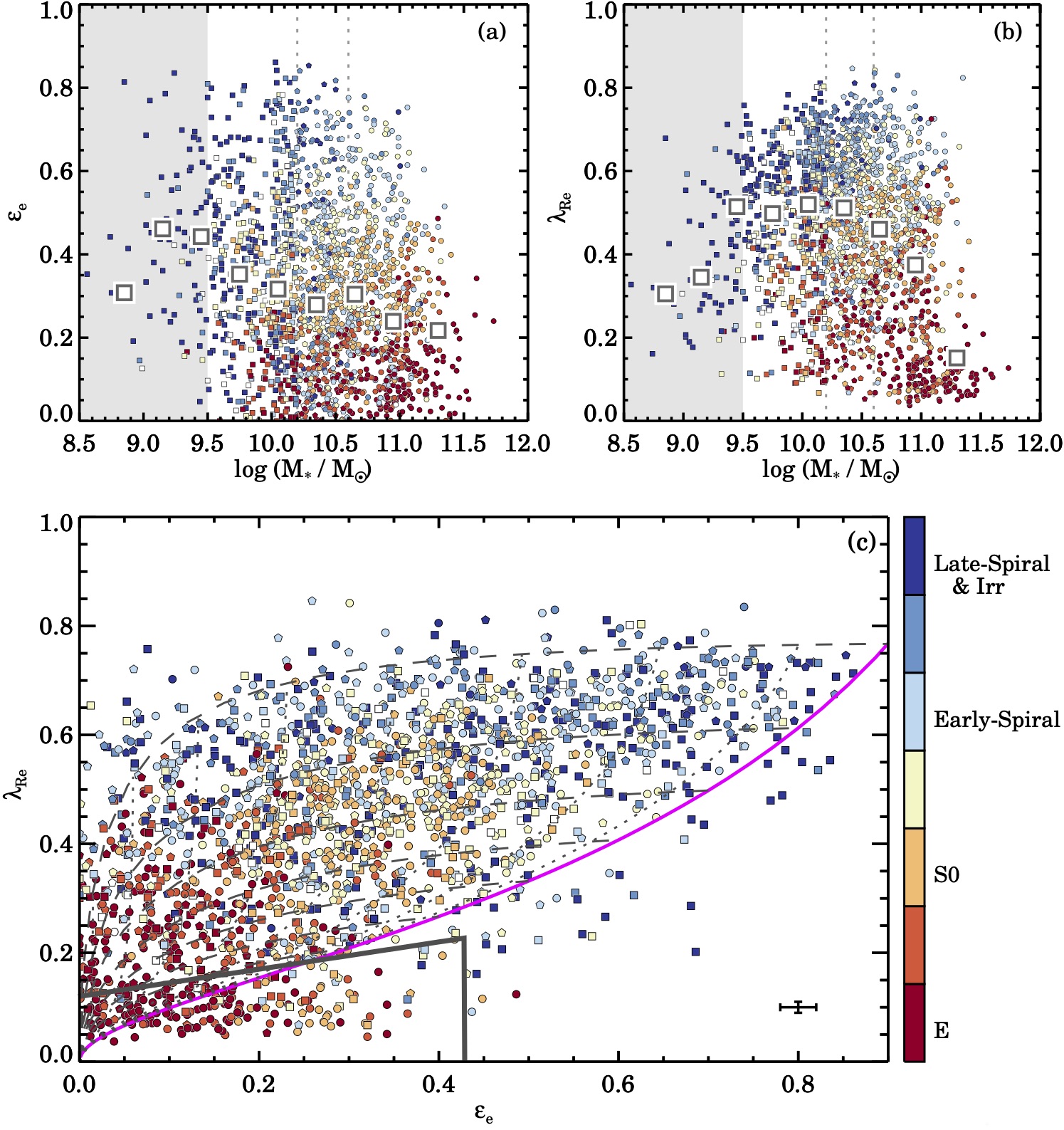}
    \caption{Stellar kinematic sample presented in the ellipticity \ee\ versus stellar mass diagram (Panel a), seeing corrected spin parameter proxy \lre\ versus stellar mass (Panel b), and \lre\ versus \ee\ (Panel c). Data are colour coded by the visual morphological type, where red-yellow colours indicate an early-type morphology, blueish colours show late-type galaxies, whereas unfilled symbols indicate that no conclusive morphology could be determined. Galaxies below $\logm<9.5$ are not used in the main analysis, but are shown in Panel (a-c) for completeness. Above $\logm>9.5$ we split the sample into three tertiles in stellar mass, indicated by the vertical lines. We use a different symbols for each mass-bin; from low- to high-mass we show individual galaxies as squares, pentagons, and circles, respectively. The larger open squares show the median of the observed sample in mass bins of width 0.3 dex. 
In Panel (a) we find that with increasing stellar mass, galaxies on average become rounder. Elliptical galaxies have the lowest mean \lre\ values and spiral galaxies have the highest mean \lre\ (Panel b). We show \lre\ versus $\ee$ in Panel (c), where we indicate theoretical predictions for the edge-on view of axisymmetric galaxies with anisotropy $\beta_z = 0.7\times \varepsilon_{\rm{intr}}$ as the magenta line \citep{cappellari2007,emsellem2011} assuming $\kappa=0.97$. Galaxies with different intrinsic ellipticities are shown by the dashed lines, going from $\varepsilon_{\rm{intr}}$=0.3 (bottom) to $\varepsilon_{\rm{intr}}$=0.9 (top). The dotted lines show the model galaxies with different viewing angle from edge-on (magenta line) to face-on (towards zero ellipticity). The solid lines on the bottom-left shows the SR selection box from \citet{vandesande2021}. The median uncertainty for the sample is shown in the lower right of panel (c).}
    \label{fig:mass_ellip_lambdar}
\end{figure*}


\section{Dynamical properties of galaxies as a function of stellar mass and environment}
\label{sec:dyn_prop}

\subsection{Mass and Morphological Dependence of the Kinematic Sample}
\label{subsec:mass_morpho}

We present the structural and kinematic measurements of our sample in Fig.~\ref{fig:mass_ellip_lambdar}. This figure sets up the framework adopted throughout this paper. In order to control for stellar mass when investigating the impact of environment, we start by dividing the sample into three stellar mass bins, such that we have approximately equal numbers of galaxies in the three bins throughout the paper: $9.5<\logm\leq10.2$, $10.2<\logm\leq10.6$, and $\logm>10.6$. In Fig.~\ref{fig:mass_ellip_lambdar}a we find that the lowest-mass bin is dominated by late-type spirals, whereas the highest stellar mass bin contains mostly ellipticals, S0s, and early-spirals. The most homogeneous distribution in galaxy morphology --- with roughly equal numbers of Es and S0s and spirals --- is found within the medium mass bin. Related to the morphological bias with stellar mass, we also find a strong dependence of \ee\ on stellar mass. On average galaxies become rounder with increasing stellar mass \citep[see also][]{kelvin2014}, although the distribution is flat between $10.2<\logm<10.6$.

In Fig.~\ref{fig:mass_ellip_lambdar}b, we show the seeing-corrected spin parameter proxy \lre\ as a function of stellar mass. Within the lowest-mass bin we find a strong increase in \lre\ from $\logm\sim9.0$ to $\logm\sim9.75$, in particular for late-type spirals. While SAMI's adopted spectral resolution limit for measuring velocity dispersions is $\sim35\kms$, and surface brightness limits could lead to a biased sample in this low stellar mass regime, we note that a similar result was found by \cite{falconbarroso2015} for galaxies in the CALIFA survey. They suggested that the low \lre\ values for these galaxies might be due the presence of a relatively large dark matter halo that could support a dynamically hot but geometrically thin stellar disk. The lowest-mass bin also contains very few galaxies with $\lre<0.1$. Between $10.2<\logm<10.6$, we find only a minor  
decline of \lre\ with stellar mass, while there is strong decrease in \lre\ at $\logm>11$, with a high density of galaxies at $\lre< 0.1$. This trend is also (qualitatively) reproduced by cosmological simulations of galaxy formation \citep[e.g.][]{penoyre2017,schulze2018,lagos2018b,choi2018,vandesande2019,walomartin2020}.

We show the combination of \lre\ and \ee\ in Fig.~\ref{fig:mass_ellip_lambdar}c that is now commonly used to dynamically classify galaxies as fast and slow rotators (see Section \ref{subsec:fsr}). Because of the \ee\ and \lre\ trends with stellar mass, the average properties of galaxies from different mass bins occupy different regions of the \lre- \ee\ space. For example, galaxies in the most massive bin on average have lower \lre\ and \ee\ as compared to galaxies in the medium mass bin. This highlights the fact that we need to control for stellar mass when studying the impact of environment on the dynamical properties of galaxies. 

We find a relatively large fraction of low-mass galaxies on the RHS of the magenta line that are inconsistent with being simple axisymmetric, rotating oblate spheroids. As shown by \citet{emsellem2011} and \citet{krajnovic2011}, these are likely to be galaxies with kinematically-decoupled cores or counter-rotating disks or bulges or late types with lower stellar mass. Within the SR selection region (solid black line), we find ten galaxies that are visually classified as late-type or irregular; the majority (7/10) with stellar mass below \logm<10.1. We note that with the deeper and higher spatial resolution Subaru-Hyper Suprime Camera DR1 imaging \citep{aihara2018}, as compared to the SDSS imaging on which the visual classification was performed, 7/10 are clear face-on spirals (some with strong bars), or have irregular morphology (1/10), with the remaining two having unclear morphology. These galaxies are excluded from the selected SR sample described in Section \ref{subsec:fsr}.

In Figs.~\ref{fig:mass_ellip_lambdar}a-c, we illustrate the impact of stellar mass on the dynamical properties in our sample. In the results that follow, we will take two different approaches to control for stellar mass. First, we will look at the fraction of slow rotators independently as a function of stellar mass and environment. The second approach will be to investigate the kinematic properties of all galaxies using the continuous intrinsic \lr\ measurements for individual galaxies as derived in Section \ref{subsubsec:lredgeon} and investigate trends with stellar mass and environment independently.

\begin{figure*}
	\includegraphics[width=0.95\linewidth]{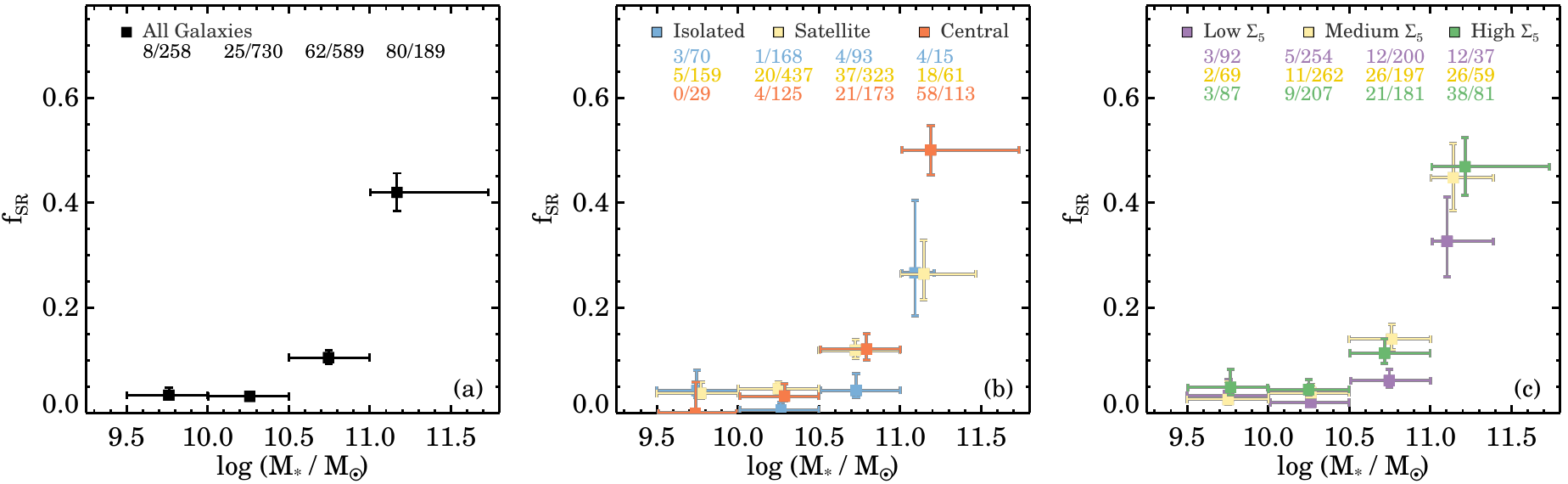}
    \vspace{-0.25cm} 	
    \caption{Fraction of slow rotators as a function of stellar mass, split using different environmental proxies. We show the 16th and 84th percentiles of the binomial confidence intervals on the fractions as vertical error bars. Horizontal bars indicate the lowest and highest stellar mass galaxy in that particular bin and the central data point is the mean stellar mass. The number of SRs versus the total number of galaxies for each bin are shown on top in each panel.
The fraction of SRs increases by a factor of $\sim2$ between $9.5<\logm<11$, but we find a strong increase by a factor of $\sim4$ in the \fsr\ at $\logm>11$ for the full sample (Panel a). We show the fraction of SRs separated into isolated centrals (blue), group or cluster centrals (orange), and satellite galaxies (yellow) in Panel (b). For the majority of galaxies in our sample (between 10<\logm<11), the central and satellite samples have a higher fraction of SRs as compared to isolated galaxies. In the highest stellar mass bin, the fraction of SRs is dominated by central galaxies, whereas satellite and isolated galaxies have lower SR fractions ($3.5\sigma$ and $1.7\sigma$ below central \fsr, respectively). When using bins of different overdensity \sfive\ (Panel c) we find that above \logm>10.5, the fraction of SRs in medium and high overdensities (yellow, green) is higher than in low-density environments (purple).}
    \label{fig:fsr_mass}
\end{figure*} 

\begin{figure*}
	\includegraphics[width=\linewidth]{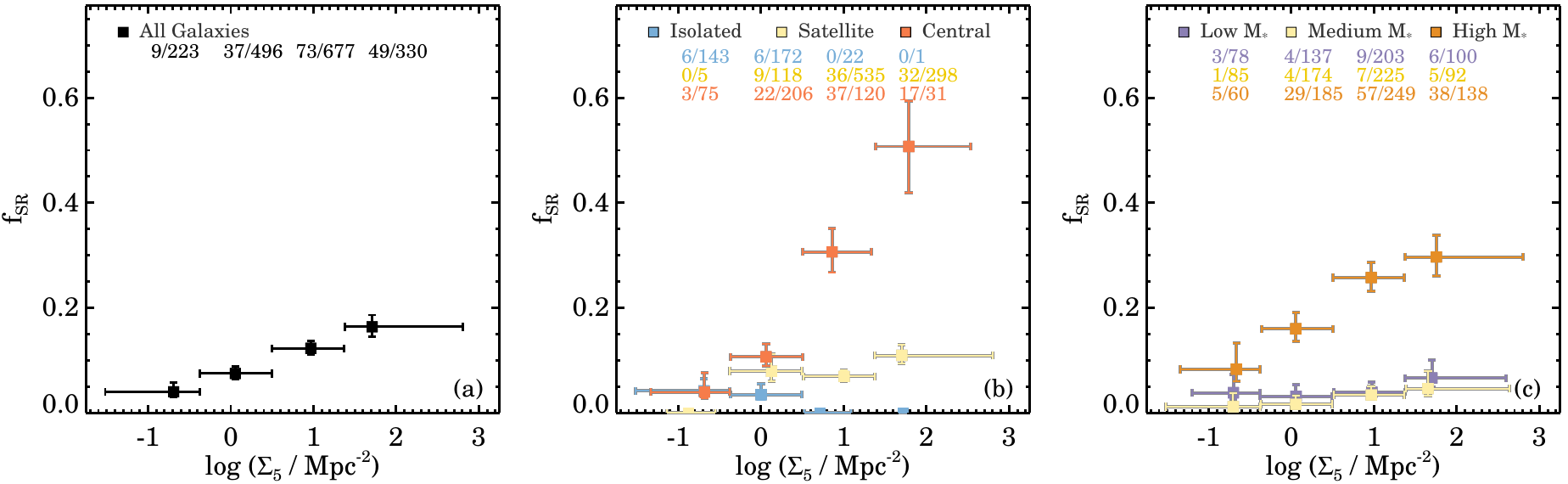}
    \vspace{-0.25cm} 	
    \caption{Fraction of slow rotators as a function of mean environmental overdensity \sfive, split by group rank and stellar mass. Figure details are the same as Fig.~\ref{fig:fsr_mass}.
The fraction of SRs steadily increases as a function of mean overdensity (Panel a). In panel (b) we find that at fixed \sfive\, group and cluster central galaxies have the highest fraction of SRs, followed by satellite galaxies. There are few SRs in isolation as explained in detail by Fig.~\ref{fig:group_vs_env}, with a low overall fraction. Panel (c) shows the sample split in three bins of stellar mass. Massive galaxies have the highest \fsr\ in all \sfive\ bins and the \fsr\ also strongly increases with \sfive, whereas we do not find a significant increase in \fsr\ as a function of \sfive\ for low and medium-mass galaxies.}
    \label{fig:fsr_fdfive}
\end{figure*} 


\begin{figure}
	\includegraphics[width=1.0\linewidth]{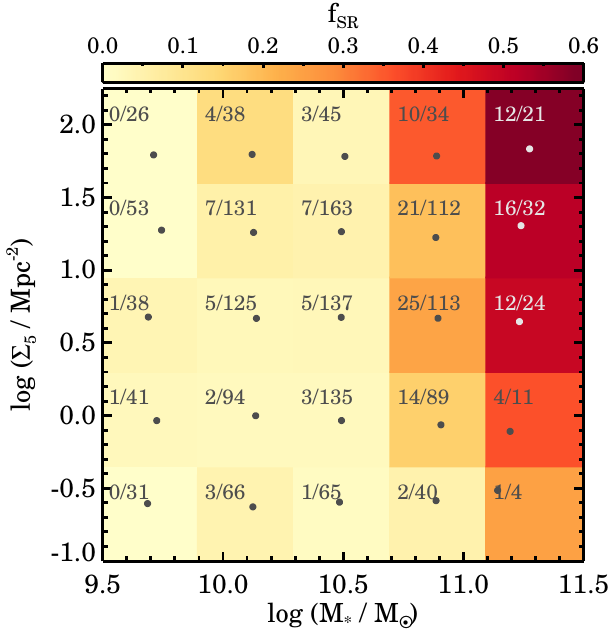}
    \vspace{-0.5cm} 	
    \caption{Fraction of slow rotators as a function of stellar mass and mean environmental overdensity. The colour coding indicates the SR fraction, whereas the numbers indicate the number of SRs to the total number in the bins. The grey dots indicate the mean stellar mass and mean environmental overdensity within each bin. At fixed stellar mass, we find that only the two highest mass bins shows a clear increase in the SR fraction with mean environmental overdensity, whereas at fixed stellar \sfive\ the \fsr\ increases as a function of mass for all galaxies.}
    \label{fig:fsr_mass_fdfive}
\end{figure}


\subsection{Fraction of SRs in different environments}
\label{subsec:fsr}

To determine the impact of environment on the kinematic nature of galaxies, we first follow the approach of previous studies and calculate the fraction of slow rotators as defined by the selection criteria from \citet{vandesande2021} that is optimised for SAMI Galaxy Survey data:
\begin{equation}
\lre < \lr_{\rm{start}} + \ee/4, \hspace*{0.1cm} \mathrm{ with } \hspace*{0.1cm} \ee < 0.35 +\frac{ \lr_{\rm{start}} }{1.538},
\end{equation}
\noindent with $\lr_{\rm{start}}=0.12$. This selection box is presented in Fig.~\ref{fig:mass_ellip_lambdar}c as the solid black line. We note that our results qualitatively do not change if we use the SR selection criteria from \citet{emsellem2011} or \citet{cappellari2016}.

We find a mean SR fraction of \fsr=$0.099 \pm 0.007$ (175 / 1766). Here, the uncertainty is calculated using the binomial confidence intervals on the fractions using the method outlined in \citet{cameron2011}. Our SR fraction is consistent with \citet{graham2018}, but lower than previous results that were based on early-type galaxy only samples \citep{emsellem2011,deugenio2013,houghton2013,scott2014,fogarty2014,brough2017,veale2017b,greene2018}. If we restrict our sample to galaxies with early-type morphology (E and S0), we find a higher fraction $\fsr = 0.172 \pm 0.012$ (175/1018). Note however, that the total number of SRs does not change if we use an early-type only sample, whereas the overall number of galaxies in the sample decreases substantially. The fraction of SRs also depends strongly on our adopted definition of early-type, which can vary depending on the depth and resolution of the image used for determining visual morphologies. Thus, we use all types of galaxies for calculating the fraction of SRs.

Compared to \citet{vandesande2017b}, who used a smaller internal release sample from the SAMI Galaxy Survey, we find a higher fraction of SRs: $\fsr=0.086\pm0.010$ versus $\fsr=0.099\pm0.007$, respectively, although the values are consistent within uncertainties. The difference between the \citet{vandesande2017b} \fsr\ and this work can be explained by the former having a larger fraction of late-type galaxies and lower median stellar mass (\logm=10.28 versus \logm=10.43, respectively). Furthermore, the current measurement also includes a seeing correction on the \lr\ measurements as well as a different SR selection region adapted for seeing-corrected data.

We present the fraction of SRs as a function of stellar mass in Fig.~\ref{fig:fsr_mass}. We calculate the SR fraction in four stellar mass bins that are equally spaced with a width of 0.5 dex between 9.5<\logm<11, whereas the highest mass bin is broader to increase the low-number statistics at the edges of the mass distribution. For each mass bin we calculate the mean stellar mass, the minimum and maximum stellar mass galaxy in that bin, the fraction of SRs, and the 16th and 84th percentiles from binomial confidence intervals on the fractions using the method outlined in \citet{cameron2011}. In Fig.~\ref{fig:fsr_mass}a we recover the well-known trend where \fsr\ increases with stellar mass. Below \logm<11\, the average SR fraction is \fsr=$0.059 \pm 0.006$ (93 / 1570), whereas above $\logm\geq11$ we find \fsr=$0.418 \pm 0.035$ (85 / 196).

In Fig.~\ref{fig:fsr_mass}b we split the sample into isolated centrals, group and cluster centrals, and satellite galaxies using the group catalogue described in Section \ref{subsec:dens_est}. For intermediate and massive galaxies (\logm>10) we find a lower fraction of isolated slow rotators as compared to group central and satellite galaxies. In each individual mass bin above \logm>10, the isolated \fsr\ is at least 1-$\sigma$ below the group central \fsr. We detect no difference between centrals and satellites, except in the highest stellar mass bin where the satellite \fsr\ is lower than the \fsr\ for central galaxies. Below \logm<10 we find mixed results between the \fsr\ of centrals, satellites, and isolated galaxies, most probably caused by the noticeably low number of SRs in these bins.

When using the local environment overdensity parameter \sfive\ (Fig.~\ref{fig:fsr_mass}c), we find similar results as for the group properties. Here, the different environment bins are defined as: $\sfiveu \leq 0.25$, $0.25< \sfiveu \leq 1.1$, and $\sfiveu>1.1$ as described in Section \ref{subsubsec:sample} and shown in Fig.~\ref{fig:group_vs_env}. Above $\logm>10$, low-density environments have the lowest fraction of SRs. Above $\logm>10.5$ the \fsr\ of galaxies in the intermediate and high \sfive\ bin are always above the \fsr\ of galaxies in the low-density bin, though with varying significance. Finally, we note that we recover the same trends if we restrict the sample to early-type galaxies only, but with larger uncertainties.

The fraction of SRs as a function of mean environmental density are presented in Fig.~\ref{fig:fsr_fdfive}, using the same approach as for Fig.~\ref{fig:fsr_mass}. We find a gradual increase of the SR fraction with environmental overdensity (Fig.~\ref{fig:fsr_fdfive}a). The relation appears to be nearly linear in strong contrast to the exponential trend between \fsr\ and stellar mass (Fig.~\ref{fig:fsr_mass}a). Split by group rank (Fig.~\ref{fig:fsr_fdfive}b), central galaxies have the highest \fsr\ that strongly increases as a function of \sfive, whereas satellites only show a mild \fsr\ increase as a function of overdensity. In Fig.~\ref{fig:fsr_fdfive}c we divide the sample into three mass bins, where only the most massive bin shows a strong increase in \fsr\ with overdensity. In principle, this strong increase in \fsr\ with \sfive\ could still be caused by the most massive galaxies residing in the highest density environments, because the highest mass bin ranges from $\logm=10.6$ to $\logm=11.75$, thus encompassing the sharp increase in \fsr\ at high stellar mass (rightmost bins in Fig.~\ref{fig:fsr_mass}a). However, if we reproduce Fig.~\ref{fig:fsr_fdfive}c including only the most-massive galaxies ($\logm>11$) in the highest-mass bin, we still observe \fsr\ to increase with \sfive\; suggesting that the observed increase is not caused by mass alone.

To investigate this further, we simultaneously show the fraction of SRs as a function of mass and environment in Fig.~\ref{fig:fsr_mass_fdfive}. By spreading the data into a larger number of bins, the typical uncertainty in each bin increases to $\pm0.05$, with extreme outliers in the lower-right bin that has an uncertainty of $\pm0.22$ due to the low number statistics. Fig.~\ref{fig:fsr_mass_fdfive} demonstrates clearly that at fixed environmental density (horizontally) the strong increase in SR fraction occurs at $\logm>11$. However, at fixed stellar mass (vertically) we still detect an increase in \fsr\ for galaxies with $\logm>10.5$, which is particularly clear for galaxies with $\logm>11$. 

Within the highest stellar mass bins, we detect a mean 0.11\,dex increase in the stellar mass from the lowest overdensity bin ($\logm=11.15$) to the highest overdensity bin ($\logm=11.26$). Because the fraction of SRs changes rapidly within this mass range, even a small change in mean stellar mass could potentially mimic a trend with environment. To investigate this possible bias, we go back to Fig.~\ref{fig:fsr_mass}a, where we find that the fraction of SRs increases by a factor of $\sim3$ over 0.5 dex in stellar mass (upper limit). As such, a 0.1 dex increase in the mean stellar mass can only explain part of the trend with environment in the highest mass bins. Thus, we conclude that at fixed stellar mass, there is an increase in the SR fraction as a function of mean environmental overdensity.

\begin{figure*}
	\includegraphics[width=\linewidth]{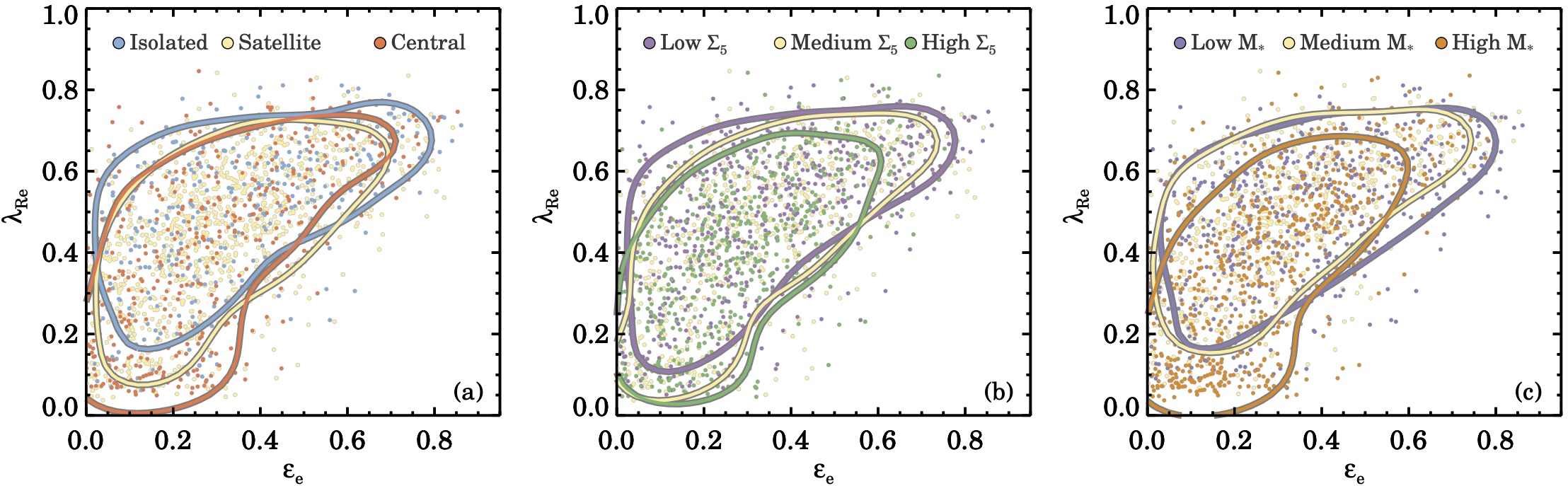}
    \vspace{-0.5cm} 	
    \caption{Seeing-corrected spin parameter proxy \lre\ versus ellipticity \ee, colour coded by group rank (central, satellite, or isolated; Panel a), mean overdensity \sfive\ (Panel b), and stellar mass (Panel c). The contours enclose 68 per cent of the total probability using kernel density estimates. Isolated centrals, satellite, and group and cluster central galaxies diverge in the low \lre\ regions, but in particular isolated galaxies extend to higher \lre\ and \ee. For the three \sfive\ bins we see the largest difference between low and high \sfive\, where galaxies in low and intermediate-density environments on average are more likely found at higher \lre\ and \ee. Galaxies in the highest stellar mass bin dominate towards low \lre\ and \ee, whereas galaxies in the low and intermediate stellar mass bins occupy the same region.}
    \label{fig:lr_intr_bins}
\end{figure*}

\subsection{The Independent Impact of Mass and Environment on Galaxy Dynamics for All Galaxy Types}
\label{subsec:lr_eo_environment}

Selecting slow rotators has been fruitful for exploring the impact of environment on the properties of galaxies with complex dynamical structure that were likely formed through major mergers \citep[see][for a recent review on the topic]{cappellari2016}. However, as the fast and slow-rotator classification is binary, it also has its clear limitations. By analysing only SRs, we are missing the intermediate regime where galaxies potentially undergo mild dynamical changes due to their likely interactions with their environment. As more than $90$ per cent of galaxies are FRs, which exhibit a large range in \lr\ related to their intrinsic shape and dynamical structure, in the following section we therefore focus our attention on this population. 

To illustrate the variations within the FR population, in Fig.~\ref{fig:lr_intr_bins} we first present the SAMI stellar kinematic \lre-\ee\ plane split into three bins of group rank, \sfive, or stellar mass. Using kernel density estimates, we show the contours enclosing 68 per cent of the total probability of finding a galaxy from a specific group.

We find that centrals, satellites, and isolated galaxies extend to opposite outer regions (Fig.~\ref{fig:lr_intr_bins}a), where the group and cluster central galaxies have the lowest \lre\ and \ee\ values. Similarly, for the mean environmental overdensity, the difference between low and high \sfive\ is most pronounced in the top-right region of the \lre-\ee\ diagram (Fig.~\ref{fig:lr_intr_bins}b). Lastly, we find that high-mass galaxies extend towards the lowest \lre\ values, whereas low and intermediate stellar mass galaxies occupy the same region (Fig.~\ref{fig:lr_intr_bins}c). This relatively simple comparison of the mass and environment bins indicates that both parameters are related to the properties of fast-rotator galaxies.

Based on these results, we will now use the inclination-corrected or intrinsic \lrei\ estimates from Section \ref{subsubsec:lredgeon} to quantify the kinematic variation of galaxies with different stellar masses in a range of environments \footnote{A similar approach was taken in \citet{brough2017} who used $\lre/\sqrt{\ee}$ as an approximate correction on \lr\ for the effects of inclination on their sample of early-type galaxies. However, as we show in Appendix their method is not well adapted for a sample with a broad range in morphology and intrinsic shape.}. We now also exclude all slow rotators and a few remaining galaxies with $\lrei<0.15$ from our sample. 

\begin{figure*}
	\includegraphics[width=\linewidth]{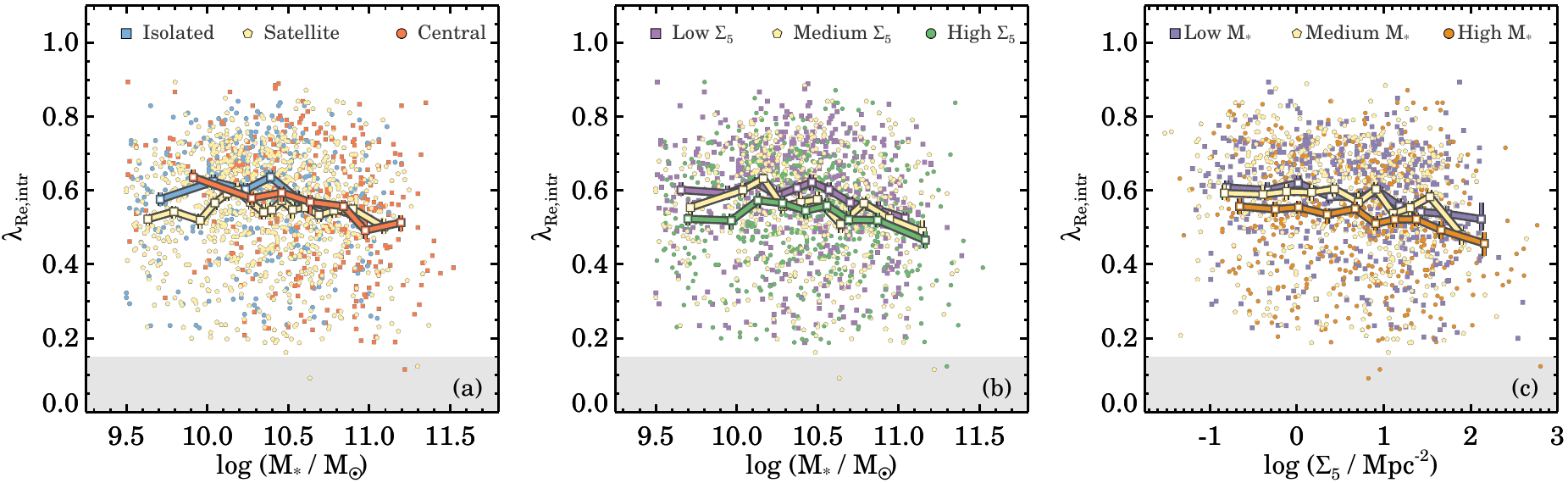}
    \vspace{-0.25cm} 	
    \caption{Intrinsic \lre\ versus stellar mass and mean environmental overdensity. Data are colour coded according to their separation in group rank, mean overdensity \sfive, and stellar mass. The solid lines with square symbols on top show the mean \lrei\ in bins that contain $\sim50$ galaxies, where the (small) vertical error bars are derived from bootstrapping the distribution. Only fast-rotator galaxies and galaxies with $\lrei>0.15$ are included here. At fixed stellar mass, we find different trends in \lrei\ with a galaxy's group property and mean local environment. Independent of stellar mass, satellites (Panel a) or galaxies in high-density environments (Panel b), on average have marginally lower intrinsic \lre\ ($\sim0.05$) as compared to isolated and low \sfive\ galaxies. However, for the three bins of stellar mass as a function of \sfive\ (Panel c) we detect a trend similar in magnitude as seen in Panel (b).}
    \label{fig:lr_intr_mass}
\end{figure*} 

\begin{figure*}
	\includegraphics[width=\linewidth]{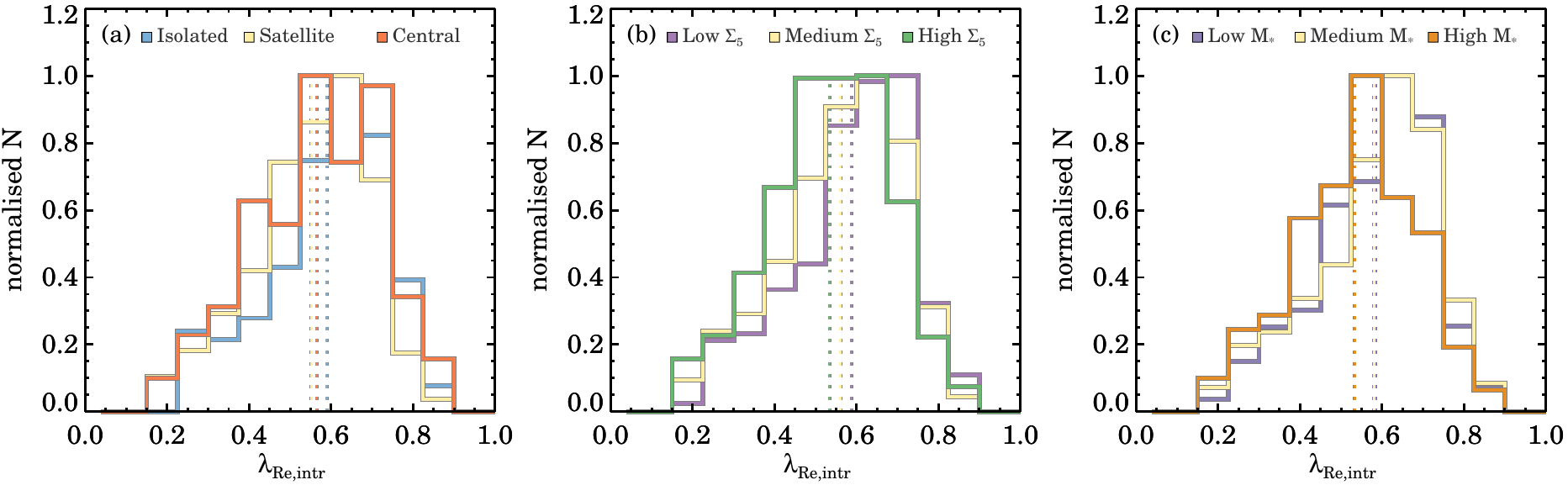}
	\vspace{-0.25cm} 
    \caption{Normalised distribution of \lrei\ for fast-rotating galaxies. Distributions are colour coded according to their separation in group properties, mean overdensity \sfive, and stellar mass. The dotted lines show the mean \lrei, with typical uncertainties on the mean \lrei\ around $\pm$0.01 derived from bootstrapping the sample. In Panel (a) we find that the distribution of satellites and group or cluster central galaxies are more skewed towards low values of \lrei\ than isolated central galaxies. Similarly, in Panel (b), the galaxies in high-density environments have the lowest \lrei\ values, followed closely by galaxies in medium and low-density environments. The difference between the stellar mass distributions is similarly small (Panel c), with galaxies in the highest mass bin having the lowest \lrei.}
    \label{fig:lr_intr_distr}
\end{figure*}

We present these measurements in Fig.~\ref{fig:lr_intr_mass} where we show \lrei\ as a function of stellar mass (Fig.~\ref{fig:lr_intr_mass}a-b) and mean overdensity (Fig.~\ref{fig:lr_intr_mass}c). In each panel, we separate the sample into three groups. The solid lines in Fig.~\ref{fig:lr_intr_mass} show the mean \lrei\ in relatively small bins of stellar mass, where each bin contains a minimum of 50 galaxies. We estimate the uncertainty on the mean values by bootstrapping the data in each bin and remeasuring the mean, and then reiterating that process 1000 times. The 16th and 84th percentile of that distribution are shown as error bars in Fig.~\ref{fig:lr_intr_mass} (note that these are relatively small and therefore hard to see).

We find that satellite galaxies (Fig.~\ref{fig:lr_intr_mass}a) have the lowest \lrei\ up to $\logm<10.75$, closely followed by central galaxies, whereas isolated galaxies are spinning the fastest. Above $\logm>10.75$ the differences between satellite, central, and isolated galaxies are insignificant. For the different lines of mean environmental overdensity as shown in Fig.~\ref{fig:lr_intr_mass}b, we also detect a trend in \lrei\ with environment. Galaxies in low-density environments on average have higher \lrei\ ($\sim0.05$) as compared to galaxies in high-density environment, with the intermediate density curve in between. Fig.~\ref{fig:lr_intr_mass}c shows that with increasing mean overdensity \sfive, the mean \lrei\ declines. The overlap of the curves for low and intermediate stellar mass, closely followed by the high-mass line, demonstrates that the trend in Fig.~\ref{fig:lr_intr_mass}b is not caused by stellar mass. Thus, our results show that both environment and stellar mass play a significant role in determining the kinematic properties of regular rotating galaxies.

To explore the trend with environment further, we show the distribution functions of \lrei\ split by stellar mass and mean overdensity in Fig.~\ref{fig:lr_intr_distr}. The \lrei\ distribution split for galaxies with different group ranking in Fig.~\ref{fig:lr_intr_distr}a shows a trend where satellite galaxies have a marginally lower intrinsic spin than central and isolated galaxies. The peak of the satellite distribution is offset from central and isolated galaxies, with a mean for each distribution of $0.552\pm{0.011}$, $0.566\pm{0.008}$, and $0.589\pm{0.008}$, respectively. We also find that the overall distribution of satellite galaxies is more skewed towards the lowest \lrei\ values as compared to isolated galaxies. This is confirmed at a $>2-\sigma$ level using a Kolmogorov–Smirnov (KS) test where the probability that isolated-centrals and satellites are drawn from the same \lrei\ distribution is $3.0\times10^{-5}$, and for isolated and centrals $3.6\times10^{-2}$.

In Fig.~\ref{fig:lr_intr_distr}b we find that galaxies in the highest environment overdensity bin have the lowest \lrei, with the mean $\lrei=0.534\pm{0.016}$ as compared to $0.561\pm{0.012}$, and $0.588\pm{0.007}$ for the medium and low overdensity bin, respectively. The overall distribution for the highest \sfive\ bin is also offset towards lower intrinsic spin. From a KS-test we also find that the probability that galaxies in low and high-densities are drawn from the same \lrei\ distribution is extremely low: $6.4\times10^{-10}$.

When comparing the different stellar mass distribution in Fig.\ref{fig:lr_intr_distr}c, the difference between the three mass bins is similar as compared to group-ranking or mean overdensity. The peak of the highest stellar mass bin is offset towards lower intrinsic spin, with mean \lrei\ of the highest and lowest mass bin differing by $-0.049\pm 0.016$ ($0.533\pm{0.011}$ versus $0.581\pm{0.012}$, respectively). Based on a KS test, there is a high probability that low and intermediate mass galaxies are drawn from the same \lrei\ distribution (0.97), whereas for low and high-mass galaxies this is extremely unlikely ($4.0\times10^{-9}$).

\begin{figure*}
	\includegraphics[width=1.0\linewidth]{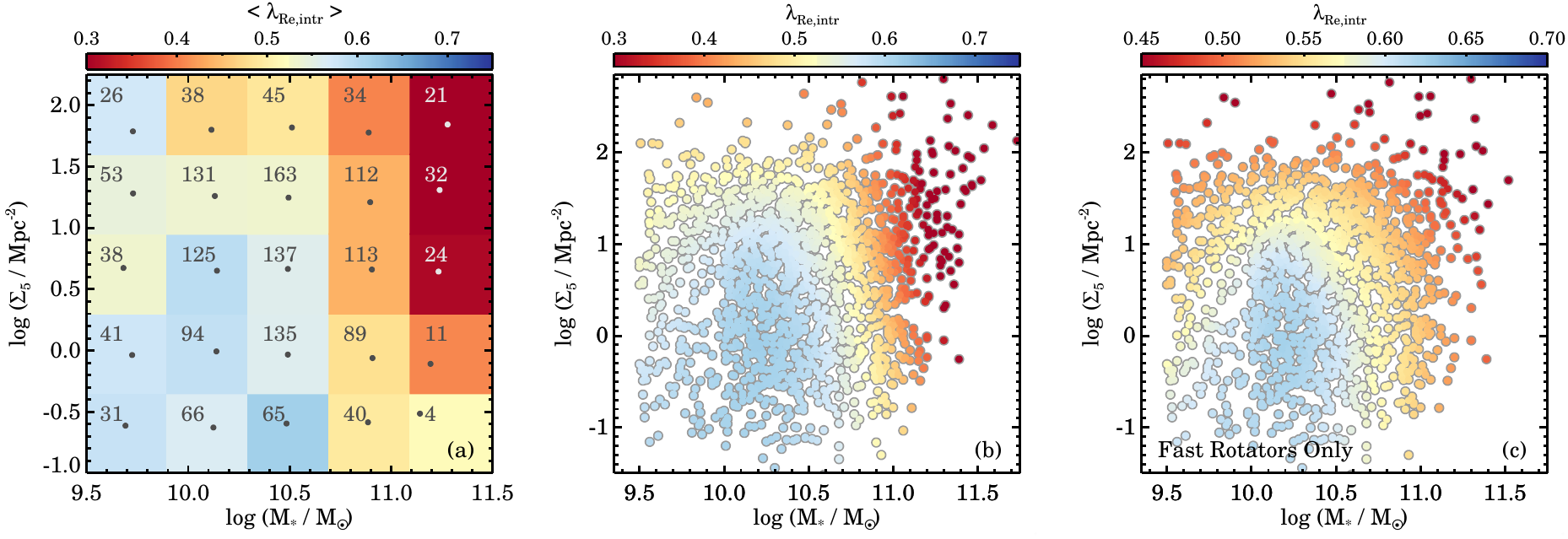}
    \vspace{-0.5cm} 
    \caption{Environmental overdensity versus stellar mass, colour coded by the intrinsic \lr. In Panel (a) we show the data in four bins of stellar mass and \sfive. The numbers indicate the total number of galaxies in each bin, and the grey dots show the mean stellar mass and mean environmental overdensity within each bin. The data for all individual galaxies are shown in Panel (b) whereas in Panel (c) we exclude slow rotators. In Panel b-c we use the LOESS smoothing algorithm to recover the mean underlying trend in \lrei. For the sample including SRs (Panel a-b), we find that at fixed environment, above $\logm>10.5$, \lrei\ decreases rapidly with increasing stellar mass. Similarly, above $\logm>10.5$ we find that \lrei\ also decreases towards higher-density environments. When excluding SRs (Panel c) the independent impact of mass and environment on the spin parameter proxy becomes clear for FRs: at fixed stellar mass \lrei\ decreases towards high \sfive, while at fixed \sfive, \lrei\ decreases towards higher stellar mass with equal strength. {We find the same results when using the observed \lre\ values (see Fig.~\ref{fig:app_lr_intr_mass_fdfive}), indicating that the inclination correction does not impact the detected trend between environment, mass, and galaxy dynamics.}
}
\label{fig:lr_intr_mass_fdfive}
\end{figure*}

Next, we simultaneously investigate galaxies as a function of mass and environmental overdensity. In Fig.~\ref{fig:lr_intr_mass_fdfive}a we first use a binning approach similar to Fig.~\ref{fig:fsr_mass_fdfive}, whereas Fig.~\ref{fig:lr_intr_mass_fdfive}b-c shows the individual data adopting a locally weighted regression algorithm \citep[LOESS; ][]{cappellari2013b} to recover the mean underlying trend in \lrei\ in a similar fashion as \cite{peng2010} and \citet{mcdermid2015}. Note that in Fig.~\ref{fig:lr_intr_mass_fdfive}a-b we analyse the complete stellar kinematic sample including SRs, whereas Fig.~\ref{fig:lr_intr_mass_fdfive}c only FRs are included. The colour coding of the data is such that red indicates a lower \lrei\ and blue a higher \lrei. The typical uncertainty on the mean across all bins in Fig.~\ref{fig:lr_intr_mass_fdfive}a is $\lrei=0.023$ with the largest uncertainty on the top-right bin (high-mass, high-density) of $\lrei=0.117$.  

We see a trend in mean intrinsic \lr\ with environmental overdensity in all but the lowest stellar mass bin in Fig.~\ref{fig:lr_intr_mass_fdfive}a: at fixed stellar mass galaxies in high-density environments have lower \lrei. The same is more clearly seen in Fig.~\ref{fig:lr_intr_mass_fdfive}b, where the LOESS smoothing reveals two independent trends between \lrei\ and stellar mass as well as \lrei\ and mean environmental density. Nonetheless, the trend in \lrei\ with stellar mass is stronger most likely due to the impact of SRs towards higher stellar mass. However, in Fig.~\ref{fig:lr_intr_mass_fdfive}c, where we have excluded slow rotators from the sample, we find that the same trend exist for fast rotators too, albeit with a smaller amplitude. Similar to what we found in Fig.~\ref{fig:lr_intr_mass}, the impact of mass and environment on \lrei\ is of similar order: from low to high \sfive, as well as from low to high \logm, \lrei\ decreases by $\sim0.15$ {with typical random uncertainties on \lrei\ of $\sim0.025$ and potential systematic uncertainties of \lrei<0.03).}

{Below \logm<10 in Fig.~\ref{fig:lr_intr_mass_fdfive}b-c, we detect a small decrease in the LOESS recovered \lrei\ values as compared to $10<\logm<10.5$. We investigate whether this trend could be caused by our halo mass correction, as we do not apply a correction to galaxies that reside in haloes with $\loghm<14.5$ (see Sec.~\ref{subsubsec:hm_distr}). However, when we lower the halo mass correction limit down to $\loghm=12.0$ in 0.5 dex increments, we do not detect any significant change in the LOESS recovered \lrei\ distributions. Nonetheless, even though the results appear to be independent of our halo mass correction, and the fact that we also detect a similar decrease in \lrei\ towards low stellar masses in Fig.~\ref{fig:mass_ellip_lambdar}b, we consider the current sample size and observational limits towards this low stellar mass regime insufficient to draw any strong conclusions from this.}

In conclusion, by comparing the intrinsic \lr\ distributions as a function of group ranking, mean environmental overdensity, and stellar mass, we find that both stellar mass and environment are correlated with the kinematic properties of galaxies. Satellite galaxies have lower \lrei\ values than group and cluster centrals and isolated central galaxies, although the absolute differences are small $\Delta\lrei=0.04$. We find a similar trend for galaxies in different environmental overdensities with the highest versus the lowest \sfive\ bin. Equally but independently, stellar mass also correlates with \lrei\ such that with increasing stellar mass \lrei\ decreases. 

{We detect the most pronounced trend in \lrei\ as a function of mass and environment in Fig.~\ref{fig:lr_intr_mass_fdfive} using 2D binning and the LOESS algorithm. Comparing the results from Fig.~\ref{fig:lr_intr_mass_fdfive} to Figs.~\ref{fig:lr_intr_mass} \& \ref{fig:lr_intr_distr} perhaps demonstrates that analysing the broad distribution in \lr\ using averaged quantities may not be the most optimal method for detecting dynamical changes in the population as a function of stellar mass and environment. Including environment as a variable in the Bayesian mixture modelling analysis of the dynamical populations as a function of mass from \citet{vandesande2021} offers a potential way forward but is beyond the scope of this work.}
In the next section we discuss whether {the trend between mass, environment, and galaxy dynamics} is a consequence of disk fading, an early formation where a decrease in star formation activity prevented these galaxies from forming thin disks, or whether these galaxies have been through a process that made them dynamically hotter and intrinsically rounder.

\section{Discussion}
\label{sec:discussion}

\subsection{Comparison with Previous Observational Results}
The discovery of the morphology-density relation revealed that galaxy morphology depends on environment \citep{oemler1974,davis1976,dressler1980}. Given the expected dynamical differences between late-type and early-type galaxies, the existence of a kinematic morphology-density relation should not be surprising. However, as morphology and the dynamical properties of ellipticals and S0s do not correlate one-to-one \citep[e.g. the majority of ellipticals are classified as fast rotators][]{emsellem2011}, the odds of detecting this relation dynamically decreases. Nonetheless, \citet{cappellari2011b} found a higher fraction of SRs in the densest areas of the Virgo cluster as compared to lower-density environments, and subsequently this was reported by several other studies \citep{deugenio2013,houghton2013,scott2014,fogarty2014}. However, results by \citet{veale2017b}, \citet{brough2017}, and \citet{greene2017,greene2018} showed that galaxy stellar mass plays a dominant role in changing the fraction of SRs, and that after controlling for stellar mass, either no significant \citep{brough2017}, or almost no remaining correlation between environment and the \fsr\ was found \citep{veale2017b,greene2018}. \citet{graham2019b}, who adopt a hybrid method for selecting SRs from MaNGA IFS data combined with a sample of "visual" SRs candidates without kinematic measurements \citep{graham2019a}, report that a KMDR does exist at fixed stellar mass above \logm>11.3, with increasing strength towards \logm$\sim12$.

Our findings show that stellar mass correlates the strongest with \fsr, and that the correlation of \fsr\ with environment is weaker but significant. This suggests that mass is the primary driver in the formation of SRs, but that environment still plays a secondary, albeit smaller role. Independent of stellar mass, we find that the fraction of SRs increases moderately with environment: above $\logm>11$ we find that the \fsr changes from 0.33$^{+0.08}_{-0.07}$ at a mean $\sfive=-0.31$ to 0.46$^{+0.06}_{-0.05}$ at mean $\sfive=1.52$ (Fig~\ref{fig:fsr_mass}c). Our findings are consistent with those of \citet{graham2019b}, although our methods for selection SRs differ and their sample extends to higher stellar masses. 

The differences between our work as compared to \citet{veale2017b}, \citet{brough2017}, and \citet{greene2018} can be explained by a combination of sample size and range in environment. \citet{veale2017b} conclude that the kinematic morphology-density relation is driven by stellar mass, with a possible exception towards a larger \fsr\ in the highest-density environment, albeit with low statistical significance. \citet{greene2018} detect no significant difference between galaxy spin of centrals and satellites, but mention that errors in the classification of central and satellite galaxies with group finders systematically lower differences between satellite and central galaxies, at a level similar to the measurement uncertainties in their work. One of the features of the SAMI Galaxy Survey is the access to the GAMA group catalogue \cite[G3Cv1;][]{robotham2011} constructed from the deep and highly-complete GAMA spectroscopic campaign, which resulted in an improved central and satellite classification as compared to using SDSS spectroscopy alone. This, combined with a larger sample of galaxies as compared to \citet{greene2018}, could explain why we do detect a difference between the average spin of centrals and satellites above a stellar mass of $\logm>11$ (Fig. \ref{fig:lr_intr_mass}a). 

\citet{brough2017} examined the fraction of SRs within the eight SAMI clusters for a sample of early-type galaxies, whereas here we have extended our sample to cover the full range in environment from galaxies in isolation all the way to galaxies in clusters including all morphological types. As the change in slow rotator fraction as a function of environment is mild, it explains why this trend was not detected by \citet{brough2017}. If we limit Fig.~\ref{fig:fsr_mass_fdfive} to only early-type galaxies in clusters, we also do not detect a significant trend with environment. 

In a recent study, \citet{cortese2019} found that satellite galaxies also undergo little dynamical and structural change during their quenching phase. Instead, from a comparison to cosmological simulations, they found that satellites galaxies do not grow their angular momentum as fast as centrals after accreting into bigger haloes. Hence, satellites do not become dynamically hotter from environmental effects. This result is consistent with our findings of satellite galaxies having marginally lower intrinsic spin parameters \lr\ as compared to central galaxies at fixed stellar mass. 

Lastly, our findings are also supported by \citet{wang2020} who also study the kinematic-morphology of galaxies, but focus more towards its relation to the star-formation rate and stellar mass. They too detect a small environmental dependence, but only for massive galaxies ($10.9<\logm<11.5$), such that galaxies in rich groups, denser environments or group centrals have lower values of \lre. However, in contrast to our results, no environmental dependence is found for lower-mass galaxies.

\subsection{Disk Fading}

Even though recent IFS surveys, such as CALIFA \citep{sanchez2012}, the SAMI Galaxy Survey, and MaNGA include late-type galaxies in their samples, this does not necessarily imply that a mass-independent kinematic-morphology density relation should be detected easily. The dynamical properties of spiral galaxies and S0s show considerable overlap in the $\lr-\varepsilon$ space \citep{falconbarroso2015,querejeta2015,cappellari2016,vandesande2017b,graham2018}. One process that can have a considerable impact on the apparent kinematic differences between the spirals and S0s populations is disk-fading \citep[e.g.][]{croom2021b}. In this process, for a typical galaxy with an older (dispersion dominated) bulge and younger disk, a quenched and ageing stellar population can lead to decrease in the observed effective radius as the outer disk becomes less-luminous. As galaxies have increasing \lr\ profiles as a function of radius, a decrease in \re\ leads to a lower observed \lre. Whilst this process does not involve a structural or dynamical transformation, it could explain, at least partially, the trend we observe for \lrei\ to be lower towards high environmental density and stellar mass.

\begin{figure}
\begin{center}
	\includegraphics[width=0.85\linewidth]{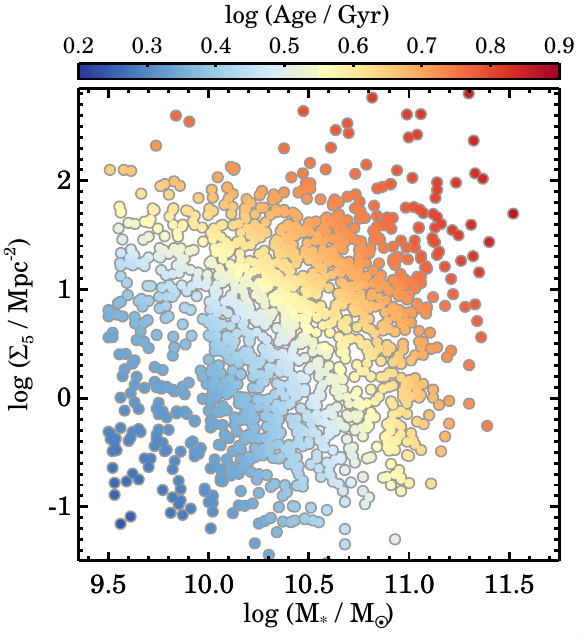}
    \caption{Environmental overdensity versus stellar mass for a sample of fast rotators, colour coded by the mean luminosity-weighted stellar population age. We use the LOESS smoothing algorithm to recover the mean underlying trend in mean $\log$(Age), where blue and red respectively indicate young and old stellar populations. We find a similar trend for age in the \sfive-\logm\ plane as seen for \lrei\ (Fig. \ref{fig:lr_intr_mass_fdfive}c). However the absolute age difference from low to high \sfive\ or \logm\ is insufficient to explain the detected trend in \lrei\ through the process of disk-fading.}
\label{fig:age_mass_fdfive}
\end{center}
\end{figure}

To test whether disk-fading, or the ageing of the galaxy's stellar population, could be cause for the change in \lrei\ that we observe for the fast-rotator population, in Fig.~\ref{fig:age_mass_fdfive} we examine the mean luminosity-weighted stellar age within one \re, in the environmental density \sfive\ - stellar mass plane \citep[for more details on the stellar populations measurements we refer to][]{scott2017,croom2021a}. We recover approximately the same trend for age as for \lrei, namely that at fixed stellar mass, from low to high environmental overdensity the mean age increases, and similarly at fixed \sfive\ the mean age increases with stellar mass \citep[see also][]{mcdermid2015}. The only difference occurs at low stellar mass (\logm<10) where we detect a small decrease in \lrei\ as compared to galaxies between $10<\logm<10.75$ (Fig. \ref{fig:lr_intr_mass_fdfive}c), whereas the mean stellar population age shows the opposite trend with galaxies becoming younger as compared to higher masses. Nonetheless, while the age and \lrei\ trends are qualitatively the same, the question is whether the absolute difference in age at low and high \sfive\ is sufficient to significantly change the observed \lrei\ values. 

Focusing on the region between $10<\logm<10.5$, we find from Fig.~\ref{fig:lr_intr_mass_fdfive}c that the largest kinematic transition occurs between $0<\sfive<2$ where \lrei\ decreases from $\sim0.65$ to $\sim0.50$ ($\Delta \lrei = 0.15$; see also Fig.~\ref{fig:lr_intr_mass}c). {Note that the typical uncertainties on \lrei\ are approximately 0.025, with possible systematic offsets less than 0.03}. At the same time, the log mean stellar age increases from $\sim0.35$ to $\sim0.65$ ($\Delta$ age = 2.25 Gyr). From table 1 in \citet{croom2021b}, we find that the largest impact of disk-fading occurs when the galaxy has just quenched. For a galaxy with bulge-to-total ratio of $B/T=0.5$ \textemdash\ where disk fading has the largest impact \textemdash\ an evolving population from 0 to 2 Gyr has a decreasing observed $\lrei=0.08$. However, for a marginally older stellar population, from 2 to 4 Gyr, \lrei\ decreases only by 0.02. Therefore, we conclude that disks-fading can only explain a minor fraction of the observed decrease in \lrei\ at different environments and stellar masses.

\subsection{Dynamical Transformation}

Several other physical processes could be responsible for the dynamical transformation of galaxies in different environments. From a theoretical perspective, many studies are aimed at explaining the formation of SRs \citep[for a review see][]{naab2014}, with minor and major mergers being the most likely, or dominant formation path. However, true slow rotators are a difficult galaxy class to reproduce \citep[e.g.][]{bendo2000,jesseit2009,bois2011}, and whether or not gas is important in the major mergers that can create slow rotators is still unclear \citep[e.g.][]{cox2006, taranu2013, naab2014, penoyre2017,schulze2018}. 

Using IFS-like galaxy mock observations from the \ea\ and \hy\ cosmological hydrodynamic observations, \citet{lagos2018b} studied the impact of environment on the formation of SRs. Their work also shows a primary dependence of the \fsr\ on stellar mass, but with a weak dependence on environment \citep[see also][]{choi2018}. \citet{lagos2018b} also detect a higher fraction of slow-rotator satellite galaxies as compared to centrals, whereas we do not find a significant difference. However, our definition of central galaxy does not include isolated galaxies in contrast to \citet{lagos2018b}. When we combine isolated and central galaxies into the same bin, we too find that satellite galaxies have a higher fraction of SRs at fixed mass. While \citet{greene2018} do not detect a significant \fsr\ difference of central and satellite galaxies at fixed stellar mass, we note that their sample is based on early-type galaxies only, and their stellar mass region of overlap between satellites and centrals is relatively small.

The formation of SRs in \ea\ is further investigated in \citet{lagos2020} who demonstrate that in most SRs, quenching (primarily due to AGN feedback) occurs before kinematic transformation, with the exception of SR satellite galaxies where these processes happen simultaneously. For $\sim50$ per cent of satellite SRs, environment played a crucial role in the kinematic transformation due to interactions with the tidal field of the halo and the group central galaxy, whereas the remaining satellite SRs were created in satellite-satellite mergers. Most importantly, in most SRs quenching seemed to be crucial for the merging process to efficiently lower the ratio of ordered to random motions.

For the dynamical evolution of fast-rotator galaxies, both gas stripping and dynamical interaction could play a dominant role. Given that the total masses of the groups in our sample are not extremely high (median $\loghm=12.9$), dynamical interactions through mergers and flybys seem the most likely scenario for making galaxies dynamically hotter and rounder \citep[e.g.][]{bekki1998,querejeta2015}. For example, \citet{bekki2011} show that repetitive slow encounters of spiral galaxies with other galaxies within a group, can transform those galaxies into S0s with thicker disks, consistent with what we find here. However, in dense environments the relative velocities between galaxies become larger, the interactions are faster, and mergers occur less frequently. Thus the interactions in clusters are less disruptive relative to those occurring in lower mass groups. As we see the impact of environment on the kinematic properties of galaxies in medium and dense environments, dynamical interactions alone cannot explain our findings.

An alternative scenario to merger-driven dynamical evolution is that galaxies at high redshift are naturally thicker and dynamically hotter, whereas a combination of continuous gas accretion, star formation and lack of strong interactions is the only way to form a colder thin disk. This scenario is proposed by \citet{lagos2017} who use the EAGLE cosmological simulations to study the specific angular momentum of galaxies ($j_\star$). They find a strong relation between $j_\star$ and both star formation and merger history. The two distinct scenarios that are proposed to give rise to galaxies that by $z=0$ have modest specific angular momentum are 1) galaxy mergers, and 2) the early quenching of star formation. While the first scenario is connected to forming slow rotators due to mergers of different mass ratio, the second scenario can link the intrinsic \lr\ evolution to environment.

This alternative scenario also naturally connects the relation between high-redshift galaxy gas disks that have been shown to be geometrically thick \citep{tacconi2013} with larger random motions \citep{forsterschreiber2009} than disks in the present-day Universe. Rotating galaxies at $z\sim2.3$ also have higher ionised H$\alpha$ gas velocity dispersions as compared to $z\sim0.9$ \citep{wisnioski2015}. As H$\alpha$ is a direct tracer of star formation, this suggests that newly formed stellar populations at lower redshifts are more likely to reside in colder rotating disks. 

A relation between disk thickness and age is also seen in our Milky Way \citep[for a recent review see][]{blandhawthorn2016}, where the thick disk is older \citep{bensby2014} with higher stellar velocity dispersion \citep{sharma2014} than the thin disk that has lower velocity dispersion \citep{lee2011}. Furthermore, on average, galaxies with older ages also have dynamically thicker disks or lower intrinsic ellipticity \citep{vandesande2018}. {In light of this, the trend in Fig.~\ref{fig:age_mass_fdfive} between in environment, stellar mass, and mean stellar population age therefore suggests that this trend should be accompanied by thicker disks in high-density environments. When we replace \lrei\ in Fig.~\ref{fig:lr_intr_mass_fdfive}c with \epsi\ we indeed find that this is the case: at fixed stellar mass, towards denser environments \epsi\ decreases. Similarly, when keeping environment as a fixed parameter, \epsi\ decreases towards high stellar mass.}

Environmental processes that can quench the star formation in these galaxies, e.g. halting gas accretion and the removal of cold gas \citep[starvation and stripping; for a recent review see][]{cortese2021} play a crucial role in this scenario \citep[see also][]{owers2019}. If star formation is ceased due to the group and cluster environment early-on, this also stops the formation of a colder, thinner component. Hence, we expect to find galaxies with thicker and hotter disks in high-density environments, because their gas accretion and star formation were truncated by the environment.

In summary, the results presented here support a scenario where slow rotators are predominantly formed through a combination of major and minor mergers. Yet, the perceived dynamical evolution within a population of more typical galaxies (e.g. spirals and S0s) can happen through early environmental quenching and the inability for galaxies to reform a cold disk combined with more recent interactions that can dynamically heat and thicken the disk. 

\section{Conclusions}
\label{sec:conclusion}

The role of environment in the formation and evolution of galaxies has been the focus of many studies since the discovery that the fraction of late and early-type galaxies changes as a function of environmental density. Whether the change in morphological properties is accompanied by a change in the structural and dynamical properties remains unclear. The key to answering this question is the availability of a large sample of galaxies with resolved kinematic measurement covering a wide range of morphology, stellar mass, in different environments. This study is now possible due the recent wealth of data from multi-object integral field spectroscopic surveys.

Using data from the SAMI Galaxy Survey, we present a study on the impact of environment on the dynamical properties for all types of galaxies. Because galaxy stellar mass and environment are closely related, the main challenge and key goal of this paper is to separate the relative importance and independence of both parameters on galaxy dynamics. To do so, we have measured the kinematic properties of 1831 galaxies across the full range in visual morphology from early to late type, with stellar masses between $9.5<\logm<12$, and a wide range of environments from galaxies in large groups and clusters to galaxies in isolation. Our key results are:

(i) \textbf{Stellar mass is the dominant driver in the formation of slow rotators}. Following a similar approach to previous studies, we begin by dividing our sample into fast rotators (FRs) and slow rotators (SRs) using the spin parameter proxy \lre\ and ellipticity \ee\ combined with the selection criteria from \citet{vandesande2021}. We find that the mean fraction of SRs (\fsr) strongly increases above a critical mass of $\logm\sim~11$ ($\fsr=0.42\pm\ 0.035$, Fig.~\ref{fig:fsr_mass}a). Across the full range in stellar mass and environment we find a mean fraction of SRs (\fsr) is $0.099 \pm 0.007$ (175 / 1766). Below $\logm<11$, the fraction of SRs is relatively low ($\fsr=0.059\pm\ 0.006$) and only increases slowly up to a stellar mass of \logm$\sim11$.

(ii) \textbf{Independent of stellar mass, we find a significant correlation between environment and the fraction of slow rotators}. By splitting the sample into centrals (of groups and clusters), satellites, and galaxies in isolation (i.e. centrals in low-mass haloes without a companion; Section \ref{subsec:dens_est}), we find that isolated central galaxies are less likely to be SRs as compared to centrals and satellites in groups and clusters between a stellar mass of $10<\logm<11$ (Fig.~\ref{fig:fsr_mass}a). In this mass range, we do not detect a difference in the fraction of SRs between satellite and central galaxies. However, above $\logm>11$, group and cluster central galaxies have the highest SR fraction, whereas isolated, slow-rotator central and satellite galaxies have statistically similar fractions.

We find similar results when dividing our sample into three bins of mean local overdensity as determined by the fifth-nearest neighbour parameter \sfive\ (Fig.~\ref{fig:fsr_mass}b), or when we divide the sample into three halo mass bins (Fig.~\ref{fig:app_halomass}a). Above $\logm>10.5$, the fraction of SRs is higher in the high-density environments (high halo mass) bin than in the low-density bin (low halo mass).

(iii) \textbf{Excluding slow rotators, we detect an independent but equal strength correlation between stellar mass and galaxy intrinsic spin and between environment and galaxy intrinsic spin}. To isolate the dependence of stellar mass and environment on the dynamics of galaxies, we use the inclination-corrected or intrinsic spin parameter proxy values \lrei. {By testing various inclination methods, we find that the systematic uncertainties on \lrei\ are less than 0.03, with typical random uncertainties of 0.025 (Appendix \ref{sec:app_eps_intr})}. We find that, at fixed stellar mass for fast rotators, satellite galaxies on average have the lowest intrinsic \lre\ values, closely followed by group and cluster central galaxies and isolated central galaxies (Fig.~\ref{fig:lr_intr_mass}a). Similarly, at fixed stellar mass, we find that galaxies in the highest-density environment have the lowest \lrei, while for galaxies in medium and low density environments, we have increasingly higher \lrei\ values (Fig.~\ref{fig:lr_intr_mass}b). When repeating the analysis as a function of mean overdensity, we detect a similar difference in the mean \lrei\ for galaxies in three bins of stellar mass (Fig.~\ref{fig:lr_intr_mass}c). The clearest evidence for an independent mass and environment trend with \lrei\ comes from using the locally-weighted regression algorithm LOESS (Fig.~\ref{fig:lr_intr_mass_fdfive}): at fixed stellar mass, the intrinsic \lre\ values of galaxies increase towards denser environment, and similarly, at fixed overdensity \sfive, the intrinsic \lre\ values of galaxies increase as stellar mass increases.

Our work demonstrates that both environment and stellar mass have a non-negligible impact on the dynamical nature of galaxies. While the mean differences for typical fast-rotator galaxies in different environments, with different stellar masses, are small $\lrei\sim0.05$, from the minimum to maximum environment or stellar mass, we detect an underlying trend with a strength of $\lrei\sim0.2$.

Despite our significantly larger sample of galaxies as compared to previous studies, it is clear that our analysis is only beginning to unravel the underlying mechanisms that transform galaxies. Larger statistical samples of resolved stellar-kinematic measurements in all environments pushing down towards lower stellar masses  \citep[e.g. Hector; ][]{bryant2020} will be key in taking the next step towards a detailed understanding on the impact of mass and environment on the dynamical evolution of galaxies.

\section*{Acknowledgements}

We thank the anonymous referee for the constructive comments which improved the quality of the paper. We also thank Rhea-Silvia Remus and Amelia Fraser-McKelvie for useful discussions on this topic. \\

The SAMI Galaxy Survey is based on observations made at the Anglo-Australian Telescope. The Sydney-AAO Multi-object Integral field spectrograph (SAMI) was developed jointly by the University of Sydney and the Australian Astronomical Observatory, and funded by ARC grants FF0776384 (Bland-Hawthorn) and LE130100198. The SAMI input catalogue is based on data taken from the Sloan Digital Sky Survey, the GAMA Survey and the VST ATLAS Survey. The SAMI Galaxy Survey is supported by the Australian Research Council (ARC) Centre of Excellence for All Sky Astrophysics in 3 Dimensions (ASTRO 3D), through project number CE170100013, the ARC Centre of Excellence for All-sky Astrophysics (CAASTRO), through project number CE110001020, and other participating institutions.

GAMA is a joint European-Australasian project based around a spectroscopic campaign using the Anglo-Australian Telescope. The GAMA input catalogue is based on data taken from the Sloan Digital Sky Survey and the UKIRT Infrared Deep Sky Survey. Complementary imaging of the GAMA regions is being obtained by a number of independent survey programs including GALEX MIS, VST KiDS, VISTA VIKING, WISE, Herschel-ATLAS, GMRT and ASKAP providing UV to radio coverage. GAMA is funded by the STFC (UK), the ARC (Australia), the AAO, and the participating institutions. The GAMA website is: http://www.gama-survey.org/.

JvdS acknowledges support of an ARC Discovery Early Career Research Award (project number DE200100461) funded by the Australian Government and funding from Bland-Hawthorn's ARC Laureate Fellowship (FL140100278). 
JBH is supported by an ARC Laureate Fellowship FL140100278. The SAMI instrument was funded by Bland-Hawthorn's former Federation Fellowship FF0776384, an ARC LIEF grant LE130100198 (PI Bland-Hawthorn) and funding from the Anglo-Australian Observatory. 
LC is the recipient of an ARC Future Fellowship (FT180100066) funded by the Australian Government. 
FDE acknowledges funding through the H2020 ERC Consolidator Grant 683184. 
NS acknowledges support of an ARC Discovery Early Career Research Award (project number DE190100375) funded by the Australian Government and a University of Sydney Postdoctoral Research Fellowship. 
SB acknowledges funding support from the ARC through a Future Fellowship (FT140101166). 
JJB acknowledges support of an ARC Future Fellowship (FT180100231). 
RMcD is the recipient of  an ARC Future Fellowship (project number FT150100333).
M.S.O. acknowledges the funding support from the ARC through a Future Fellowship (FT140100255). 
AMM acknowledges support from the National Science Foundation under Grant No. 2009416. 
Parts of this research were conducted by ASTRO 3D, through project number CE170100013. 

This paper made use of the \texttt{mpfit} \textsc{IDL} package \citep{mpfit}, as well as the \texttt{astropy} \textsc{python} package \citep{astropy}, \texttt{IPython} \citep{ipython}, the \texttt{matplotlib} plotting software \citep{matplotlib}, the scientific libraries \texttt{numpy} \citep{numpy}, and \texttt{scipy} \citep{scipy}. 


\section*{Data availability}

All observational data presented in this paper are available from Astronomical Optics' Data Central service at https://datacentral.org.au/ as part of the SAMI Galaxy Survey Data Release 3.

\clearpage

\bibliographystyle{mnras}
\bibliography{jvds_sami_kmdr_arxiv}


\appendix

\section{Using halo mass as a proxy for environment}
\label{sec:halo_mass}

\begin{figure*}
\includegraphics[width=0.32\linewidth]{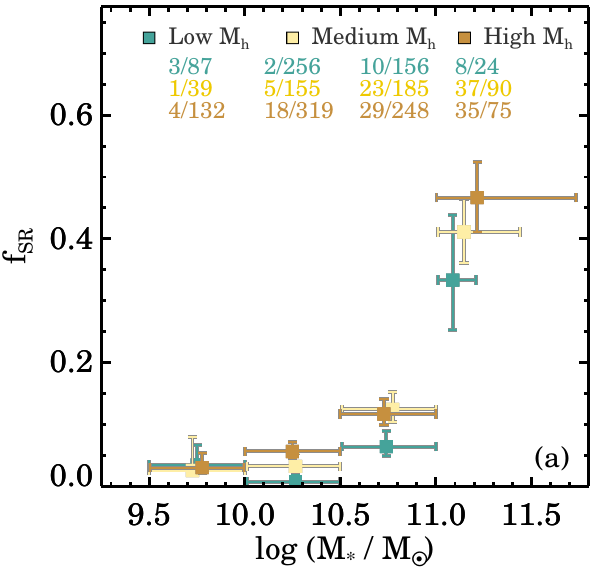}
\includegraphics[width=0.32\linewidth]{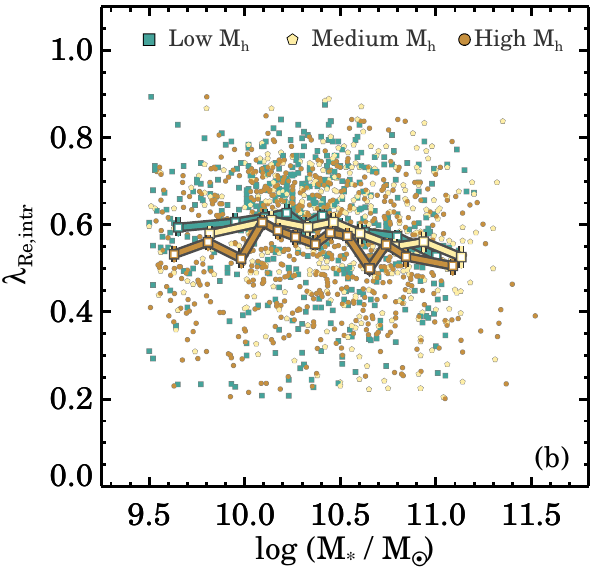}
\includegraphics[width=0.327\linewidth]{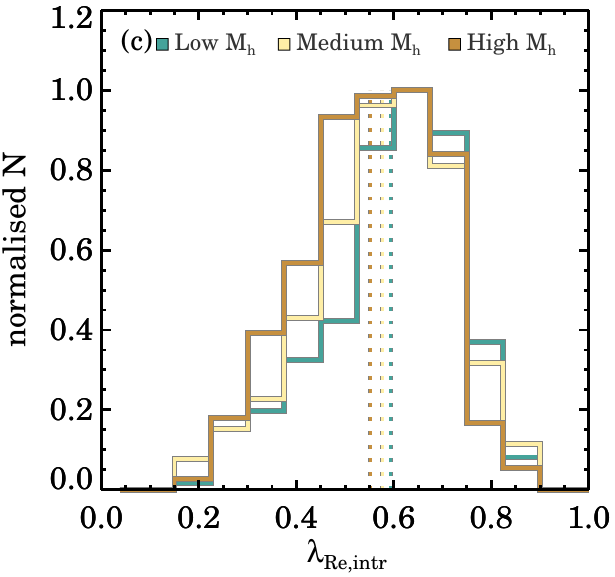}
\caption{Repeating the analysis of Section \ref{subsec:fsr} and \ref{subsec:lr_eo_environment} but using halo mass to separate galaxies into three different bins of environment. In Panel (a) we find similar results to Fig.~\ref{fig:fsr_mass}c, albeit with lower statistical significance: galaxies in low mass haloes have lower slow rotator fractions between $10.5<\logm<11.0$. In Panel (b) and (c) we confirm the results of Fig.~\ref{fig:lr_intr_mass} and Fig.~\ref{fig:lr_intr_distr}: galaxies in high-mass haloes have lower \lrei.}
\label{fig:app_halomass}
\end{figure*}

In this Section, we repeat the analysis of Section \ref{subsec:fsr} and \ref{subsec:lr_eo_environment} using the group and cluster halo masses to separate the samples into three different environmental bins. The following three limits for the low, medium, and high halo mass bins are adopted: \loghm<12.5 (which included isolated galaxies), 12.5<\loghm<14, \loghm>14. The results and main conclusions are presented in Fig.~\ref{fig:app_halomass}.

\section{Approximating Edge-on Spin Values using $\lr$ and $\varepsilon$}
\label{sec:app_eps_intr}

\subsection{Testing inclination correction methods}
\label{sec:app_incl_method_test}

\begin{figure*}
\includegraphics[width=\linewidth]{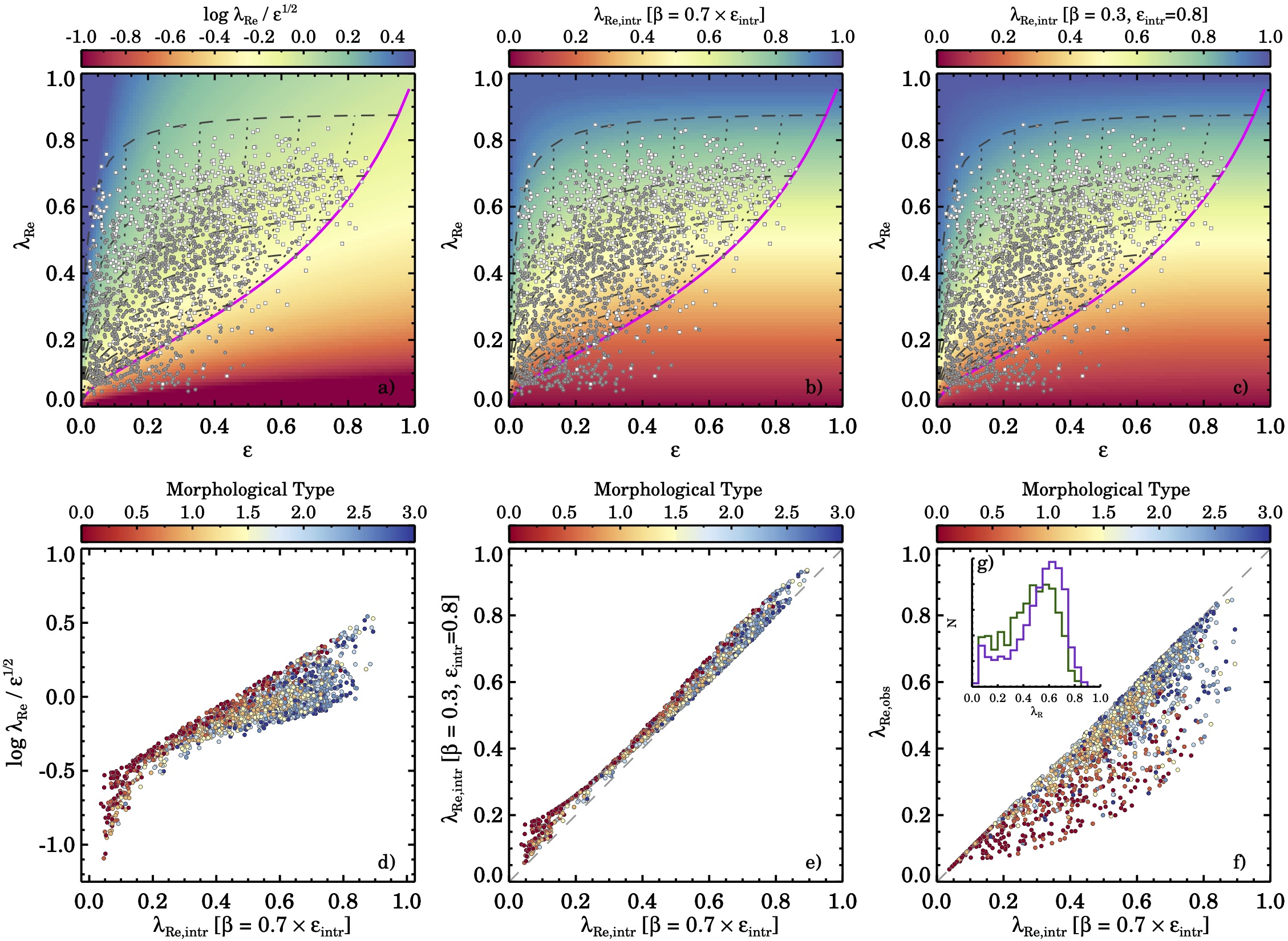}
\caption{Comparison of the approximation for the inclination corrected \lre\ values. In Panel (a)-(c) the background colour reflects the inclination corrected \lre\ values from \citet{brough2017}, this work, and \citet{frasermckelvie2021}. Furthermore, we show tracks of \epsi\ and inclination as described in Section \ref{subsubsec:lredgeon}, as well as SAMI Galaxy Survey data split into early-types (grey circles) and late-types (white squares). 
The inclination correction in Panel (b) and (c) agree well across the full range in \lre\ and \ee\, whereas the inclination correction in Panel (a) shows different a stronger divergence at low \lre\ (dark-red colour) and at \lre>0.6. Panel (d) presents a quantitative comparison of the method from \citet{brough2017} (as seen in Panel a) and the method from Section \ref{subsubsec:lredgeon} (as seen in Panel b). For the vast majority of early-type galaxies (red-orange) both methods agree reasonably well, but at $\lrei<0.2$ as well as for late-type galaxies there is considerable scatter. Our default estimated \lrei\ values agree well with the estimates from \citet{frasermckelvie2021} that were optimised for late-type galaxies (Panel e). { In Panel (f) we compare the observed \lre\ measurements to the \lrei\ estimates, whereas inset Panel (g) shows the distributions of \lre\ (green) and \lrei \ (purple)}.
}
\label{fig:app_intr_ellip_example}
\end{figure*}

Different methods for inclination correcting \vse\ and \lre\ exist. They vary in complexity from only using the $\sqrt{\varepsilon}$ \citep{brough2017}, to estimating the inclination whilst assuming an intrinsic ellipticity and anisotropy distribution \citep{querejeta2015,falconbarroso2019,delmoralcastro2020,frasermckelvie2021}, or using theoretical predictions \citep{binney2005} from the tensor Virial theorem that links velocity anisotropy, rotation and intrinsic shape \citep[][]{vandesande2018}, as well as fitting Jeans Anisotropic MGE models \citep{cappellari2013a,kalinova2017} or {full Schwarzschild orbit-based dynamical models \citep{cappellari2006,vandenbosch2008,walsh2012,thomas2014,zhu2018,vasiliev2020}.}

First, we compare our inclination corrected spin parameter from Section \ref{subsubsec:lredgeon} to the method employed by \citet{brough2017} and \citet{frasermckelvie2021}. Our goal is to investigate how these different inclination corrections behave in different regions of the $\lr-\varepsilon$ diagrams. This will be valuable for comparing the results from samples that consisted of early-type galaxies only \citet[e.g.][]{brough2017}, or star-forming galaxies \citep{frasermckelvie2021}, or all morphological types as used here, because early-type galaxies typically occupy the lower-left area of the \lr-\ee\ diagram, whereas late-type galaxies fill the upper half of the \lr\-\ee\ diagram.

\cite{brough2017} use the $\lr/\sqrt{\varepsilon}$ as an approximation for the inclination corrected \lre\ for early-type galaxies. This correction is demonstrated in Fig.~\ref{fig:app_intr_ellip_example}a, where we show the $\lr-\varepsilon$ diagram with the background colour indicating what value of $\log(\lr/\sqrt{\varepsilon})$ a galaxy in that region would have. Superimposed, we show the various model tracks for axisymmetric galaxies (see Fig.~\ref{fig:mass_ellip_lambdar}) combined with the sample that we use throughout this paper split in early-types (grey circles) and late-types (white squares). For the galaxy sample below \lre<0.5, which mostly consists of early-type galaxies, we find that $\log(\lr/\sqrt{\varepsilon})$ traces the change in dynamical properties well as can be seen from the clear change in colour. However, above \lr>0.5, $\log(\lr/\sqrt{\varepsilon})$ varies predominantly in the $\varepsilon$ direction in disagreement with predictions from the tensor Virial theorem. 

In Fig.~\ref{fig:app_intr_ellip_example}b the colour coding shows the \lrei\ parameter derived in Section \ref{subsubsec:lredgeon}. The largest \lrei\ gradient is seen low \lre\ and \ee\, where a small change in either \lre\ or \ee\ can impact the estimated \lrei\ considerably. Below the magenta line the model has no predictive power, and this region is primarily dominated by galaxies with complex structural (e.g. triaxility) and dynamical properties (slow rotators, kinematically decoupled cores, $2\sigma$ galaxies) we assume that \lrei=\lre.

Nonetheless, these predictions that link velocity anisotropy, rotation and intrinsic shape, agree remarkably well with a more simplified version of the inclination correction (Fig.~\ref{fig:app_intr_ellip_example}c). Here we use the \lre\ deprojection as presented in \cite{emsellem2011} and \citet{frasermckelvie2021} \citep[see also][]{falconbarroso2019,delmoralcastro2020} derived from the following set of equations:

\begin{equation}
\lrei = \frac{\lre}{\sqrt{C(i)^2-\lr^2(C(i)^2-1) }}
\end{equation}

\noindent with $C(i)$:

\begin{equation}
C(i) = \frac{\sin(i)}{\sqrt{1-\beta_z \cos^2(i)}}
\end{equation}

\noindent Following \citet{frasermckelvie2021}, we assume an intrinsic axis ratio $q_0=0.2$ (or \epsi=0.8) and anisotropy value $\beta_z=0.3$, a good approximation for late-type galaxies \citep{kalinova2017}. From the colour coding alone, it is hard to detect a difference between the results from Fig.~\ref{fig:app_intr_ellip_example}b and \ref{fig:app_intr_ellip_example}c.

In Fig.~\ref{fig:app_intr_ellip_example}d, we present a one-to-one comparision of $\log(\lr/\sqrt{\varepsilon})$ and \lrei\ as presented in this paper. For galaxies with $\lrei<0.2$, the two estimates deviate from a globally linear relation. For these low \lre\ values, our method has little predictive power as most galaxies lie towards the right or below the magenta line. However, as most of these galaxies are expected to have triaxial stellar light distributions, it is unclear what the best approach for inclination correcting these galaxies is. Above $\lrei>0.2$, the scatter between the two estimates correlates strongly with morphology. For elliptical galaxies (Mtype$ < 1.0$) the scatter is small, whereas the late-type population reveals larger scatter. Thus the main conclusion from this comparison is that for an early-type only sample, both methods trace the changing dynamical properties of galaxies relatively well.

As compared to the estimates from \citet[][Fig.~\ref{fig:app_intr_ellip_example}e]{frasermckelvie2021}, we find that for late-type galaxies, both methods agree well. For all galaxies, we find a small median offset of \lrei=0.03 with a 1-$\sigma$ scatter of 0.025. The largest offset is present for early-type galaxies, which is expected given the assumptions on the intrinsic axis ratio and anisotropy value ({median offset $\lrei = 0.06$ with a $1-\sigma$ scatter of 0.02). This offset disappears if we assume an intrinsic axis ratio $q_0 = 0.8$ (or $\epsi=0.2$) and anisotropy value $\beta_z=0.2$ which are more typical for slow rotator galaxies}. 

{However, we stress that these inclination corrections are unsuitable for massive slow-rotator galaxies. For non-rotating spherical or triaxial systems $\beta_z$ is a poor estimator of the stellar orbital complexity, as $\beta_z$ expresses the anisotropy in cylindrical coordinates and converges to zero. Instead, for these types of systems, the anisotropy is typically expressed in spherical coordinates \citep[$\beta_r$; e.g. see][]{cappellari2007}. Strong radial variations of $\beta_r$ are known to occur in massive ellipticals \citep{vandenbosch2010,thomas2014} and given their kinematic misalignments \citep{franx1991} the dependence of \lr\ on the viewing angle is different from our analysis above. Nonetheless, the impact of varying inclination on the absolute change in \lr\ for slow rotators is likely small, as demonstrated in cosmological simulations \citep[e.g.][]{schulze2018}. Given that most triaxial galaxies lie outside our inclination correction region (i.e. below the magenta line), and that there is good agreement between both methods for fast rotators, we conclude that the results in this paper are not significantly impacted by the inclination correction for \lr.}

{In Fig.~\ref{fig:app_intr_ellip_example}f we compare the observed \lre\ measurements to the \lrei\ estimates, which demonstrates that the differences between both parameters are relatively small for the majority of galaxies. This is also seen in inset Fig.~\ref{fig:app_intr_ellip_example}g, where we find that the inclination correction primarily causes the peak of the \lrei\ distribution (purple) to shift to higher values as compared to the observed distribution (green).}

\begin{figure*}
\includegraphics[width=0.45\linewidth]{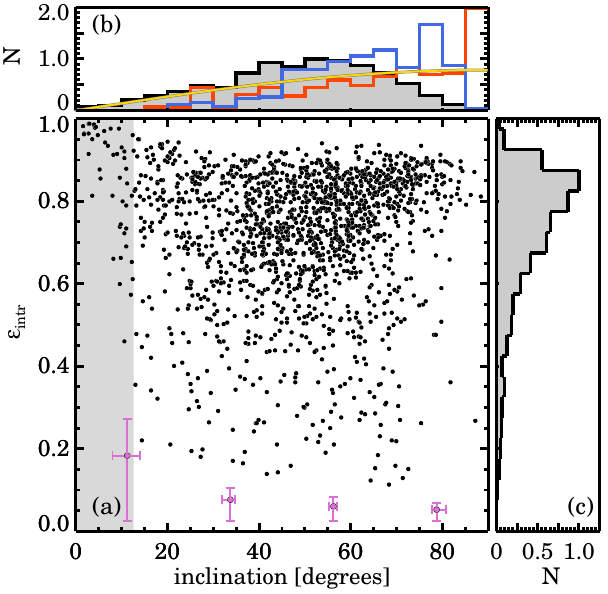}
\caption{ {Recovered intrinsic ellipticity \epsi\ and inclination $i$ derived from theoretical predictions and our data from within the observed (\lre,\ee) diagram. Only galaxies that are consistent with being axisymmetric rotating spheroids are included. Besides the individual galaxy data (black circles), in Panel (a) we also show the typical uncertainty in four inclination bins (magenta points). Panel (b) shows the recovered inclination distribution for SAMI Galaxy Survey data (black) compared to results from \at\ (red) and CALIFA (blue). The yellow line show the theoretical prediction for an intrinsically-thin disk. While the recovered inclination distribution from SAMI shows a paucity of highly-inclined galaxies, this bias does not impact the recovered intrinsic ellipticities (\epsi) or intrinsic \lre\ values. Panel (c) shows the recovered intrinsic ellipticities, but we note that this distribution is not representative of the entire stellar kinematic sample as we have only selected galaxies that lie above the magenta line in the \lre-\ee\ diagram.}}
\label{fig:app_intr_ellip_incl}
\end{figure*}

\subsection{Recovered distributions of the intrinsic ellipticity and inclination}
\label{app:epsi_incl}

{Here, we analyse the recovered distributions of the intrinsic ellipticity \epsi\ and inclination $i$. In Fig.~\ref{fig:app_intr_ellip_incl} we show all galaxies that lie above the magenta-line for which we can use the model tracks to estimate the \epsi\ and $i$. For the data presented in this figure, we did not apply a lower limit of \ee>0.025 that is invoked in the main analysis of this paper to avoid spurious results in the region where the model predictions for \lrei\ become highly degenerate. This region covered by \ee<0.025 is indicated as the gray-shaded region in Fig.~\ref{fig:app_intr_ellip_incl}a. Here, we also see that the uncertainty on the recovered \epsi\ and $i$ rise towards low inclinations, as well as a clear bias towards flattened (high \epsi) objects.}

{In Fig.~\ref{fig:app_intr_ellip_incl}b we show the distribution of inclinations from the SAMI Galaxy Survey (black) as compared to the \at\ distribution from \citet{cappellari2013a} (red) and CALIFA from \citet{kalinova2017} (blue), where the latter two are derived from Jeans Anisotropic MGE models. Furthermore, we shows the expected inclination distribution for an intrinsically thin disk observed under random viewing angles ($p(i) \propto \sin i$; yellow). As the probability for observe a thin disk edge-on is higher than face-on, the inclination distribution for all survey is skewed towards high $i$.}

{For the SAMI Galaxy Survey there is a clear drop in galaxies with recovered inclinations greater than $i>70^{\circ}$. The reason for this is two-fold. First, from Fig.~\ref{fig:app_intr_ellip_example}b we see that the area covered by the model tracks between $80<i<90^{\circ}$ is smaller than for inclination bins between $40<i<80^{\circ}$. Small errors in the observed ellipticity and \lre\ can therefore push galaxies more easily out of the $80<i<90^{\circ}$ regions as compared to other inclinations, lowering the probability that we observe galaxies towards high inclinations. Secondly, the magenta line that shows the $i=90$ projection, assumes that thinner disks (higher \epsi) have higher anisotropy following the relation $\beta_z = 0.7 \times \epsi$ \citep{cappellari2007}. However, this relation is an upper limit and regular rotators show a wide range of anisotropies \citep[e.g.,][]{cappellari2007,schulze2018}. Hence, for galaxies with lower anisotropy values than $\beta_z = 0.7 \times \epsi$ we will underestimate the inclination because they typically lie further to left of the magenta line.}

{Importantly, even though this range in possible anisotropies causes a bias in the recovered inclinations, the effect on the estimated \epsi\ and \lrei\ is marginal. From Fig.~\ref{fig:app_intr_ellip_example}b we see that the model tracks with $i>60^{\circ}$ have shallow to flat slopes in the \lre\ direction. Therefore, even if the recovered inclination is biased, the recovered \epsi\ and \lrei\ are not significantly impacted by this. In summary, we find that the results in this paper are not significantly impacted by the inclination correction for \lr.
}

\subsection{Environmental overdensity versus stellar mass without inclination corrections}

\begin{figure*}
\includegraphics[width=\linewidth]{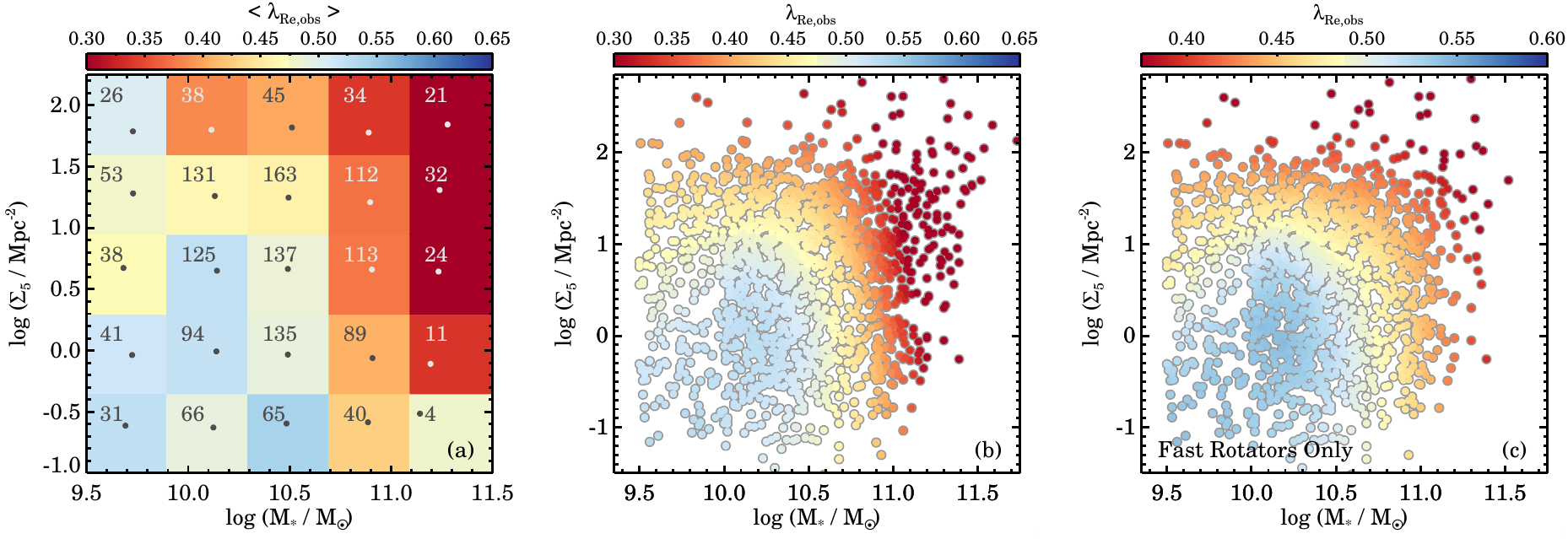}
\caption{ {Environmental overdensity versus stellar mass, colour coded by the observed \lre. This figure is similar to Fig.~\ref{fig:lr_intr_mass_fdfive} but without an inclination correction on \lre. These results demonstrate that the trend we observe between environment, mass, and galaxy dynamics is not caused by the inclination correction.}}
\label{fig:app_lr_intr_mass_fdfive}
\end{figure*}

{Lastly, to test whether our results from Section \ref{subsec:lr_eo_environment} depend on the inclination correction, here we present the equivalent of Fig.~\ref{fig:lr_intr_mass_fdfive} but now without the inclination correction on \lre. Apart from a change in the colour range for \lre, we detect no significant differences between Fig.~\ref{fig:lr_intr_mass_fdfive} and Fig.~\ref{fig:app_lr_intr_mass_fdfive}.}

\bsp	
\label{lastpage}
\end{document}